\documentclass[12pt,preprint]{emulateapj}

\shorttitle{Long time-scale behavior of the Blazhko effect}
\shortauthors{Benk\H{o} et al.}

\begin{document}

\title{Long time-scale behavior of the Blazhko effect from
rectified {\it Kepler} data}

\author{J.~M. Benk\H{o}\altaffilmark{1}, E. Plachy\altaffilmark{1,2},
R. Szab\'o\altaffilmark{1}, L. Moln\'ar\altaffilmark{1}, Z. Koll\'ath\altaffilmark{1,2}}

\altaffiltext{1}{Konkoly Observatory, MTA CSFK, Konkoly Thege Mikl\'os \'ut 15-17., H-1121 Budapest, Hungary}
\altaffiltext{2}{Institute of Mathematics and Physics, Savaria Campus, University of West Hungary \\ H-9700 Szombathely, K\'arolyi G\'asp\'ar t\'er 4, Hungary}

\email{benko@konkoly.hu}

\begin{abstract}
In order to benefit from the 4-year unprecedented precision of the 
 {\it Kepler} data, we extracted light curves from the 
pixel photometric data of the {\it Kepler} space
telescope for 15 Blazhko RR\,Lyrae stars. 
For collecting all the flux from a given target as 
accurately as possible, we defined tailor-made apertures for each star and 
quarter. In some cases the aperture finding process yielded 
sub-optimal result, because some flux have been lost even if the aperture  
contains all available pixels around the star. This fact stresses the 
importance of those methods that rely on the whole light curve instead of 
focusing on the extrema (O$-$C diagrams and other amplitude independent methods).
We carried out detailed Fourier analysis of the light curves and
the amplitude independent O$-$C diagram. 
We found 12 (80\%) multiperiodically modulated stars in our sample. 
This ratio is much higher than previously found.
Resonant coupling between radial modes, 
a recent theory to explain of the Blazhko effect, 
allows single, multiperiodic or even 
chaotic modulations. Among the stars with two modulations 
we found three stars (V355\,Lyr, V366\,Lyr and V450\,Lyr)
where one of the periods dominate in amplitude modulation, but the other period has
larger frequency modulation amplitude.
The ratio between the primary 
and secondary modulation periods is almost always very close to
ratios of small integer numbers. It may indicate the effect of undiscovered
resonances. Furthermore, we detected 
the excitation of the second radial overtone mode $f_2$ for three stars
where this feature was formerly unknown. 
Our data set comprises the longest continuous, 
most precise observations of Blazhko RR Lyrae stars
ever published. These data which is made publicly available
will be unprecedented for years to come.
\end{abstract}


\section{Introduction}

The long and (almost) un-interrupted observations of the {\it Kepler}
space telescope allow us to investigate  moderate or small amplitude
brightness variations with long periods. 
These studies generally need lots of telescope time and high precision 
simultaneously, therefore space photometry is ideal for them.

Such an interesting phenomenon is the Blazhko effect \citep{Bla07} 
of RR\,Lyrae stars, the long period amplitude (AM) and frequency 
modulation (FM) of the observed light curves. 
Formerly, the effect was defined as the presence of (at least) one of these two 
types of modulations, however, recent investigations have always found 
both of them simultaneously. The typical pulsation (light variation) period
of a fundamental mode pulsating RR\,Lyrae (RRab) star is about half a day 
with 0.5-1~mag amplitude, while the Blazhko modulation time-scale is
generally 10-1000 times longer and its amplitude is around a few tenths
of magnitudes or smaller. 

These behaviors encumbered the investigation of the Blazhko effect in the past.
Now, the long time coverage, the high duty cycle and precise observations of 
{\it Kepler} allow us to address questions such as how regular the 
Blazhko effect is and how frequent multiperiodic Blazhko stars are. 
The better knowledge of the Blazhko phenomenon
is important, because the effect is frequent 
(the incidence of the Blazhko effect among
RR\,Lyrae stars is high: 30 -- 50\% depending on the different samples 
considered) and its physical origin -- despite of the serious 
efforts during the last hundred years -- is still unknown.

As of today two competitive explanations of the Blazhko effect 
have survived among the many previously suggested ones (see for a review 
\citealt{Kol12}, or \citealt{Kov09}). Stothers' idea
\citep{Sto06} explains the effect with the influence 
of local transient magnetic fields excited by the turbulent convection. 
The first quantitative tests of this idea found 
serious inconsistencies between theory and observations 
\citep{Smolec11, Mol12}. 
\cite{Bu11} suggested a model where the modulation caused
by a resonance coupling between low order radial (typically fundamental)
mode with a high order radial (so-called strange) mode.
This model is based on the amplitude equation formalism, but
 has not been tested yet by hydro-dynamic computations. 
Both of these theories can potentially predict variable Blazhko cycles.
In the first case the variation could be 
quasi-periodic with stochastic nature, 
while in the second case it can be regular:
single or multiperodic, or chaotic.   

\section{Data}\label{data}

The main characteristics of the {\it Kepler Mission} are described in 
\cite{Koch10} and \cite{Jen10a, Jen10b}. Technical details can be
found in these handbooks: \citet{KIH,DPH,KDCH}. We summarize here  
those basic facts about the instrument only that proved to be important 
in our analysis.

\begin{figure}
\includegraphics[width=8.5cm]{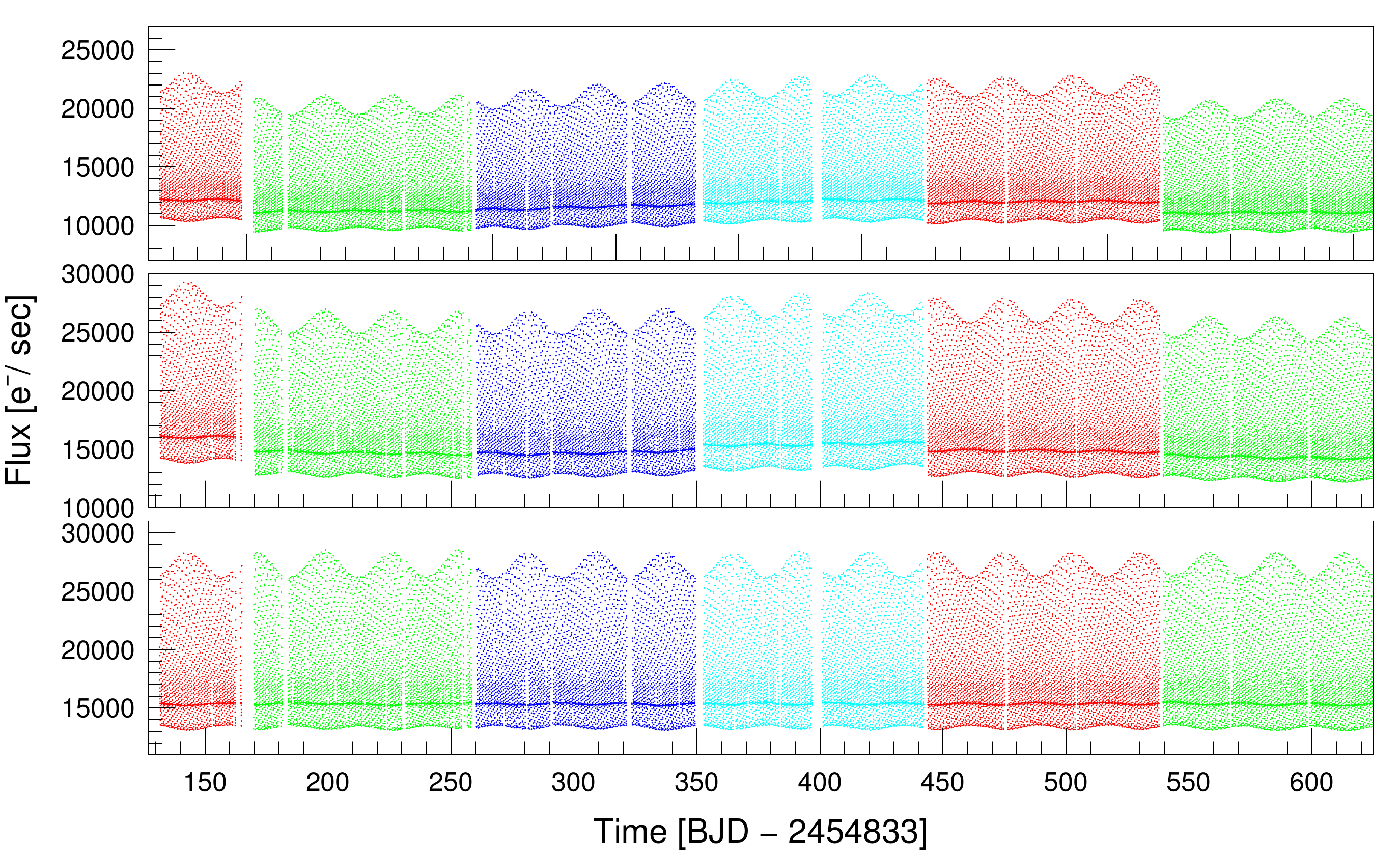}
\caption[]{
Top panel: `optimal aperture' 
flux variation curve of V783\,Cyg ({\it Kepler} archive). 
Middle panel: the flux variation curve prepared by using the best tailor-made 
aperture. Bottom panel: scaled, shifted and detrended curve. (For better visibility, 
only the first six quarters are plotted.)
} \label{llc_ny}
\end{figure}

The space telescope orbits the Sun and observed one
fixed area continuously at the Cygnus/Lyra region. To ensure the
optimal illumination for its solar cells the equipment was 
rolled by 90 degrees four times a year. 
As a consequence, each target star's light is collected on
four different CCDs according to the quarterly positions. This  
implies possible systematic differences among data from 
different quarters.
When we have a look at the flux variation curves of an RR\,Lyrae star, 
the zero points and amplitudes are 
evidently different from quarter to quarter for most of the stars 
(see Fig.~\ref{llc_ny} for an example).

Due to the limited telemetrical capacity 
only small areas around each selected target stars were downloaded.
We will refer these areas in this paper as `stamps'. Within these stamps 
`optimal apertures' (in {\it Kepler} jargon) were fitted from 
the 1024 pre-defined ones for each star and quarter separately. 
The photometry done on these apertures defines the {\it Kepler} flux variation curves. 
These apertures are, however, optimal only if the light variation of the target
is less than about of a tenth of a magnitude \citep{DPH}. 
Since the total amplitude of pulsation for {\it Kepler} RR\,Lyrae stars 
is between 0.47 and 1.1 mag \citep{Nemec13} (hereafter N13),
these pre-defined apertures are not optimal any more: significant
fraction of the flux flows out of these apertures.  
This effect adds to the fact that even 
the apertures may differ quarterly for a given star explaining 
most of the differences between the amplitudes and average fluxes
belonging to different quarters.

\subsection{The Sample}

\begin{figure*}
\includegraphics[width=8cm]{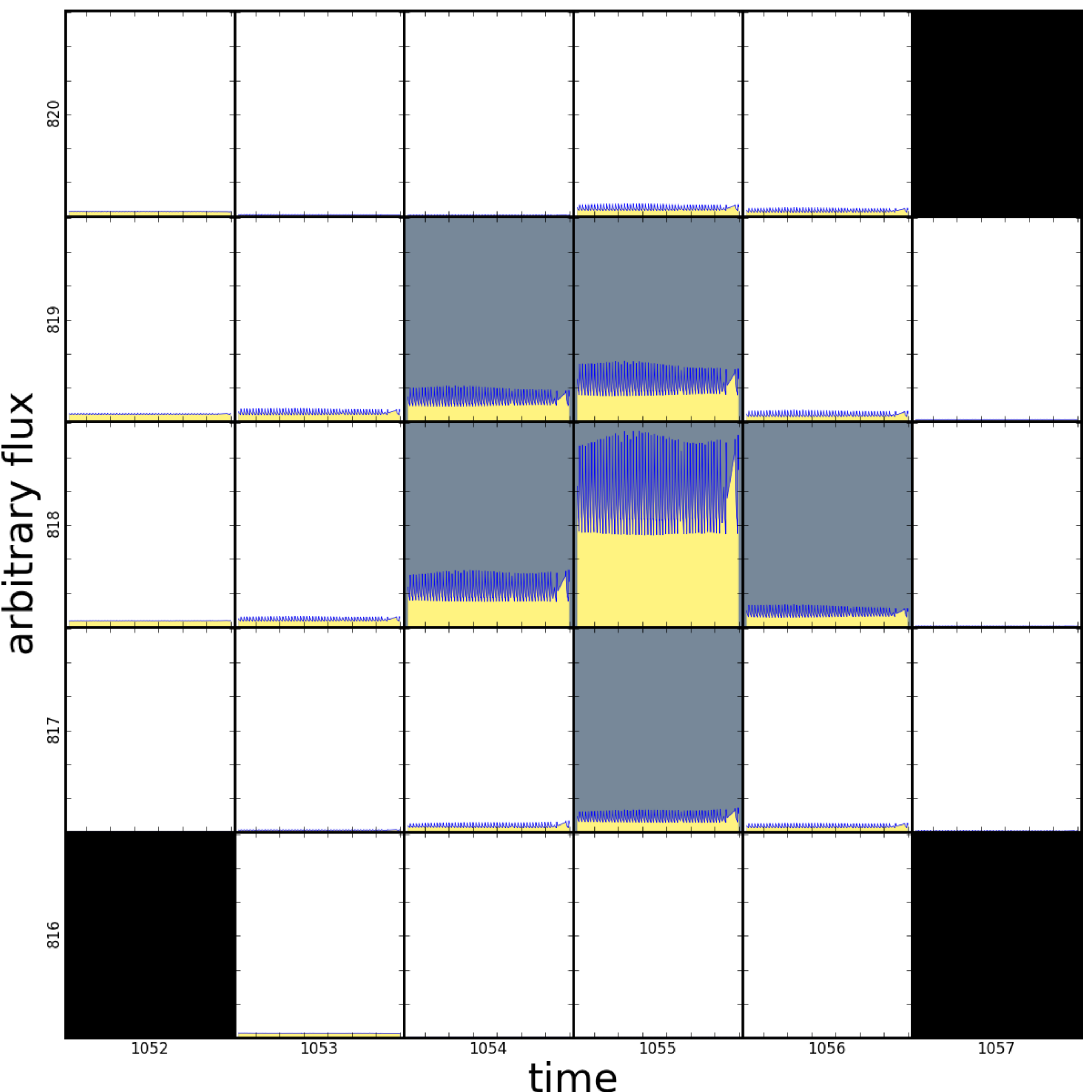}
\includegraphics[width=8cm]{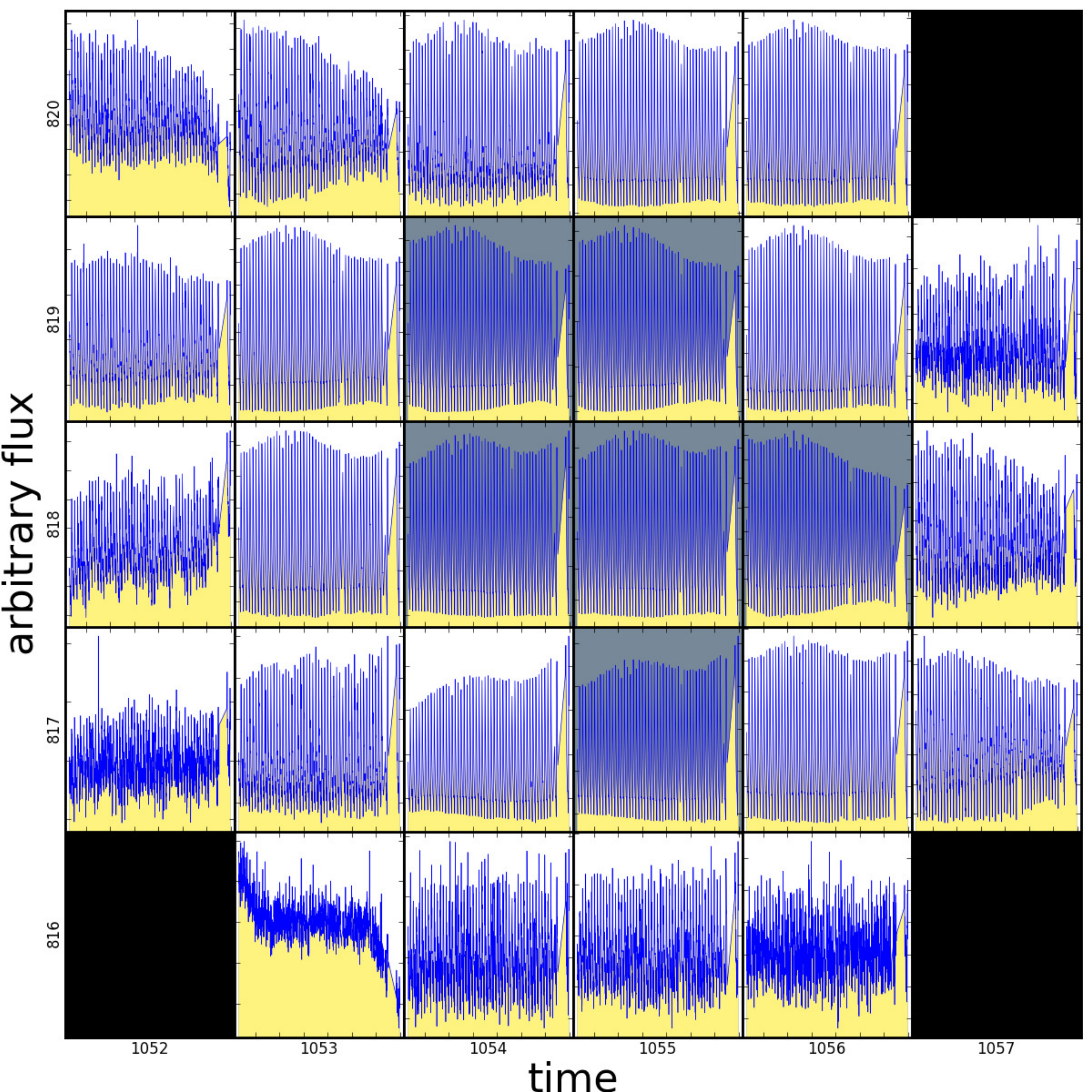}
\caption[]{
Constructing of the tailor-made apertures.
The entire pixel mask around \object{V783 Cyg} (KIC~5559631) in the
first quarter. Grey pixels: elements of the `optimal apertures', 
white pixels: all other downloaded pixels (`stamp').
We plotted the Q1 time series of each pixel individually. 
On the left hand side all pixel light curves are scaled between zero and the 
maximum flux. On the right hand side  all pixel 
light curves are scaled individually.
We collected all those pixels that show the signal of the star 
and omitted those ones 
which include noise or background sources only.
} \label{Q1_map}
\end{figure*}

We assembled our Blazhko star sample in the {\it Kepler} field the following way.
The sample of \cite{Benko10} contains fourteen stars.
One of them -- \object{V349 Lyr} (KIC~7176080) -- proved to be
a non-modulated star suggested by \cite{Nemec11}.
This finding has been confirmed by the present work by 
checking its pixel photometric data. We also include
three additional stars that were analyzed by N13.
The extremely small Blazhko effect of V838\,Cyg (KIC~10789273) 
was revealed by N13, while the (Blazhko)
RR\,Lyrae nature of \object{KIC 7257008} and \object{KIC 9973633} was discovered by 
the ASAS survey  (\citealt{Poj97, Poj02}; Szab\'o et al. in prep.)
when the {\it Kepler} measurements had been in progress. That is the reason 
why we have data on these two targets from Q10 only.
The other 13 stars were pre-selected by the {\it Kepler}
Asteroseimic Science Consortium\footnote{\url{http://astro.phys.au.dk/KASC/}} (KASC) 
and were observed during the whole mission. N13 shortly noted two additional
 Blazhko candidates, however, both of those stars are
faint ones merging with neighboring bright sources.
Because of the serious problems concerning the 
separation of their signals from their close companions
we omitted them from our investigations. (They will be discussed in a forthcoming paper.)  
\object{RR Lyr} itself is also in the {\it Kepler} field, but
its image is always saturated. Therefore, recovering its original 
signal needs extra caution and 
special techniques (e.g. custom apertures). 
Many successful efforts have been done to this direction
\citep{Kolenberg10, Kolenberg11, Molnar12}.  
We will only refer to those results on RR\,Lyr.
Our final sample consists of fifteen Blazhko stars. 

The exposition time of the {\it Kepler} camera is 6.02~s with 0.52~s readout time. 
The long cadence (LC) data result from 270 exposures co-added with a total 1766~s 
integration time. Since we concentrate on long period effects we generally 
used these LC data only, especially as the time-span of short cadence 
data (SC: 9 frames codded 58.85~s) are usually no more than one
single quarter.

The commissioning phase data (Q0) between 2009
May 2 to 11  (9.7\,d) included only one Blazhko star: \object{KIC 11125706}.
The observations of other targets began with Q1 
on 2009 May 13. Here we analyzed LC data  
to the end of the last full quarter (Q16, till 2013 Apr 8).
The total lengths of the data covers 3.9 years. 
The CCD module No.~3 failed during Q4 (on 2010 January 12), the targets 
located on this module have quarter-long gaps in their time series data.
Six stars of our fifteen-element sample suffer from this defect 
(see Tab.~\ref{Blazhko_stars} and Fig.~\ref{zoo}).
The combined number of data points for a given star is 
between 19\,249 (KIC~9973633)  and 61\,351 (V783\,Cyg),
the typical value is about 50-60\,000.
Data are public and can be downloaded from the web page of
MAST\footnote{\url{http://archive.stsci.edu/kepler/}}.

\subsection{Data Processing}\label{data_processing}

\begin{figure*}
\includegraphics[width=17cm]{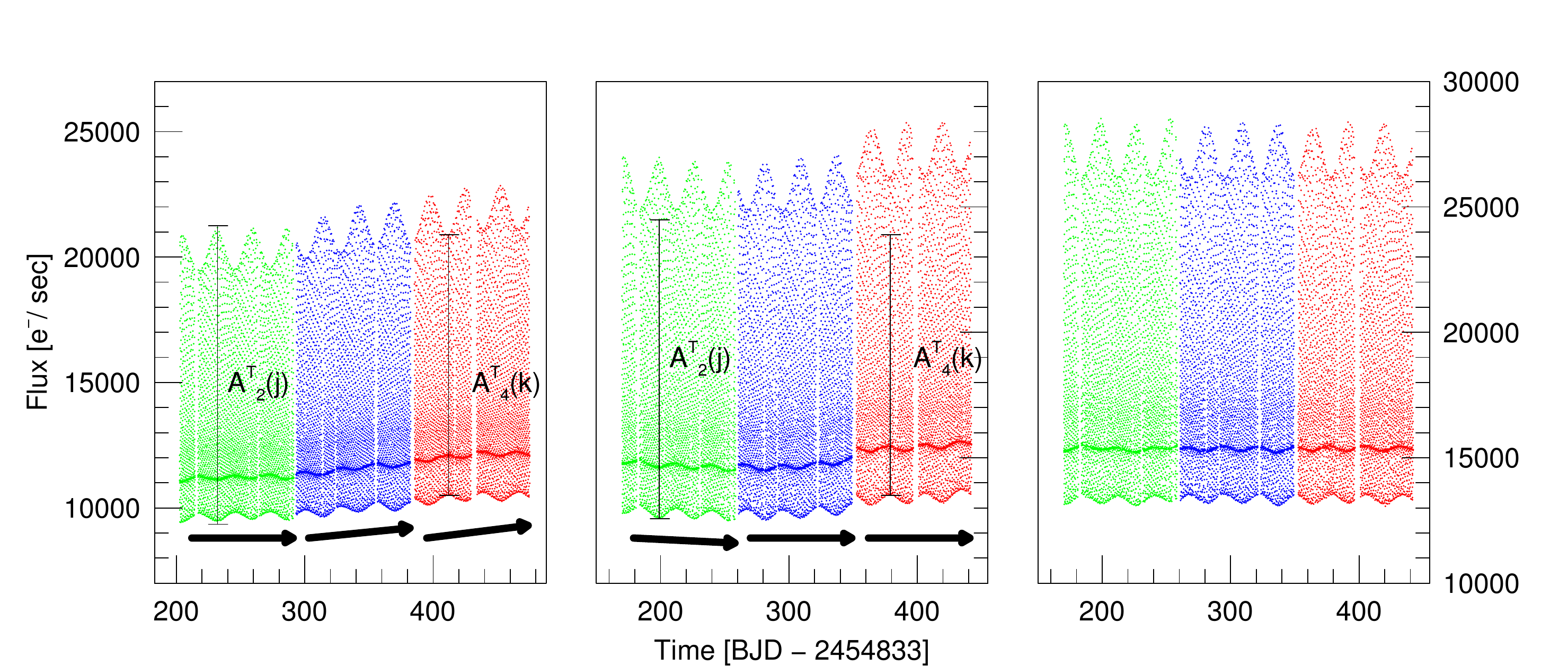}
\caption[]{
Flux variation curves from the data processing.
The figure shows curves of V783\,Cyg from Q2-Q4. 
From left to the right: the archived {\it Kepler} data,
fluxes extracted from the best tailor-made apertures, and 
final rectified data, respectively. The arrows illustrate the internal trends.
$A^{\mathrm T}_2(j)$ and $A^{\mathrm T}_4(k)$ are the total pulsation
amplitude of the archived {\it Kepler} Q2 and Q4 data at the $j$th and $k$th pulsation
cycles, respectively. These amplitudes are also shown in the tailor-made data plot for 
comparison.
} \label{llc_2}
\end{figure*}
In this subsection we summarize the main steps done before our analysis.
The {\it Kepler} data available for each source in two forms (1) as 
{\it photometric time series}: flux variation curves (flux vs. time) prepared 
from the pre-defined optimal apertures and (2) {\it image time series} 
(image of the stamp vs. time). Latter data sets are often referred to as `pixel data'. 
For the above mentioned reasons we used these
pixel data. After we downloaded them from the MAST web page
PyKE\footnote{\url{http://keplergo.arc.nasa.gov/PyKE.shtml}} 
routines provided by Kepler Guest Observer Office were used to extract 
the flux variation curves for each pixel in the stamps of a given star.
Since the pixel files before file version 5.0 have a time stamp
error we corrected it by using the PyKE {\tt keptimefix} tool. 

\begin{table*}
\begin{center}
\caption{Sample from a rectified data file}
\label{sample_data}  
\footnotesize{
\begin{tabular}{rcrccrr}
\tableline\tableline
No & Time  &   Flux  & Zero point offset
& Scaling factor &  Corrected flux &  Corrected K$_{\mathrm p}$\\  
     &   (BJD$-$2454833)          &    (e$^{-}$s$^{-1}$)         &    (e$^{-}$s$^{-1}$)   &          
& (e$^{-}$s$^{-1}$)    & (mag)  \\ 
\tableline
1 &     131.5123241 &    5322.6 &   $-400.00$  & 1.000 &     5402.03436789  & 0.39793763 \\
2 &     131.5327588 &    5393.9 &   $-400.00$  & 1.000 &     5473.33140201  & 0.38370163 \\
3 &     131.5531934 &    5496.7 &   $-400.00$  & 1.000 &     5576.12843615  & 0.36349907 \\
4 &     131.5736279 &    5498.1 &   $-400.00$  & 1.000 &     5577.52547030  & 0.36322708 \\
5 &     131.5940625 &    5488.2 &   $-400.00$  & 1.000 &     5567.62250444  & 0.36515654 \\
6 &     131.6144972 &    5571.6 &   $-400.00$  & 1.000 &     5651.01953856  & 0.34901397 \\
7 &     131.6349317 &    6347.2 &   $-400.00$  & 1.000 &     6426.61657271  & 0.20937502 \\
8 &     131.6553663 &    8478.6 &   $-400.00$  & 1.000 &     8558.01360685  & $-0.10160144$ \\
9 &     131.6758010 &    14437.0 &  $-400.00$  & 1.000 &     14516.41064097 & $-0.67531712$ \\
10 &    131.6962356 &    15817.5 &  $-400.00$  & 1.000 &     15896.90767511 & $-0.77395064$ \\
$\cdots$   &    $\cdots$   &   $\cdots$   &           $\cdots$   &      $\cdots$   &    $\cdots$   &    $\cdots$   \\
\tableline                                               
\end{tabular}
}
\tablecomments{The first ten data lines from the file of \object{V2178 Cyg} ({\tt kplr003864443.tailor-made.dat}).
The columns contain serial numbers, baricentric Julian dates, flux extracted from the tailor-made 
aperture, zero point offsets, scaling factors (1.0 = no scaling), stitched (shifted, scaled and trend filtered) flux and
their transformation into the K$_{\mathrm p}$ magnitude scale, respectively. See the text for the details.   
}
\end{center}                                            
\end{table*} 

We investigated the flux variation curves of all individual pixels separately.
We illustrate this process in Fig.~\ref{Q1_map}.
Here the stamp of Q1 around the star V783\,Cyg (KIC~5559631)
is plotted on two different scales.
The gray pixels symbolize the elements of the pre-defined `optimal' aperture.
We plotted the Q1 time series of each pixel individually. On the left panel
all pixel flux variation curves are scaled between zero and the 
maximum flux attained by the brightest pixel in the stamp. This option 
shows the relative contribution to the archived flux variation curve by each 
pixel. 
The total flux of the star comes obviously from a few pixels only.
On the right panel all individual pixel flux variation curves are 
scaled separately between their minima and maxima 
to ensure the largest dynamic range for each pixel.
This map reveals that there is some flux from outside of the
original (`optimal') aperture. 

Aiming for apertures for each star and quarter separately
 which include the total flux,
we built the `tailor-made' apertures in the following way: 
if the flux variation curve of a pixel showed the signal of the given variable star
-- that is the main pulsation period is detectable 
($A(f_0) \sim 3\sigma$) in the Fourier spectrum --
we added the pixel in question to our tailor-made aperture, 
otherwise we dropped it.  
\begin{figure*}
\includegraphics[width=18.5cm]{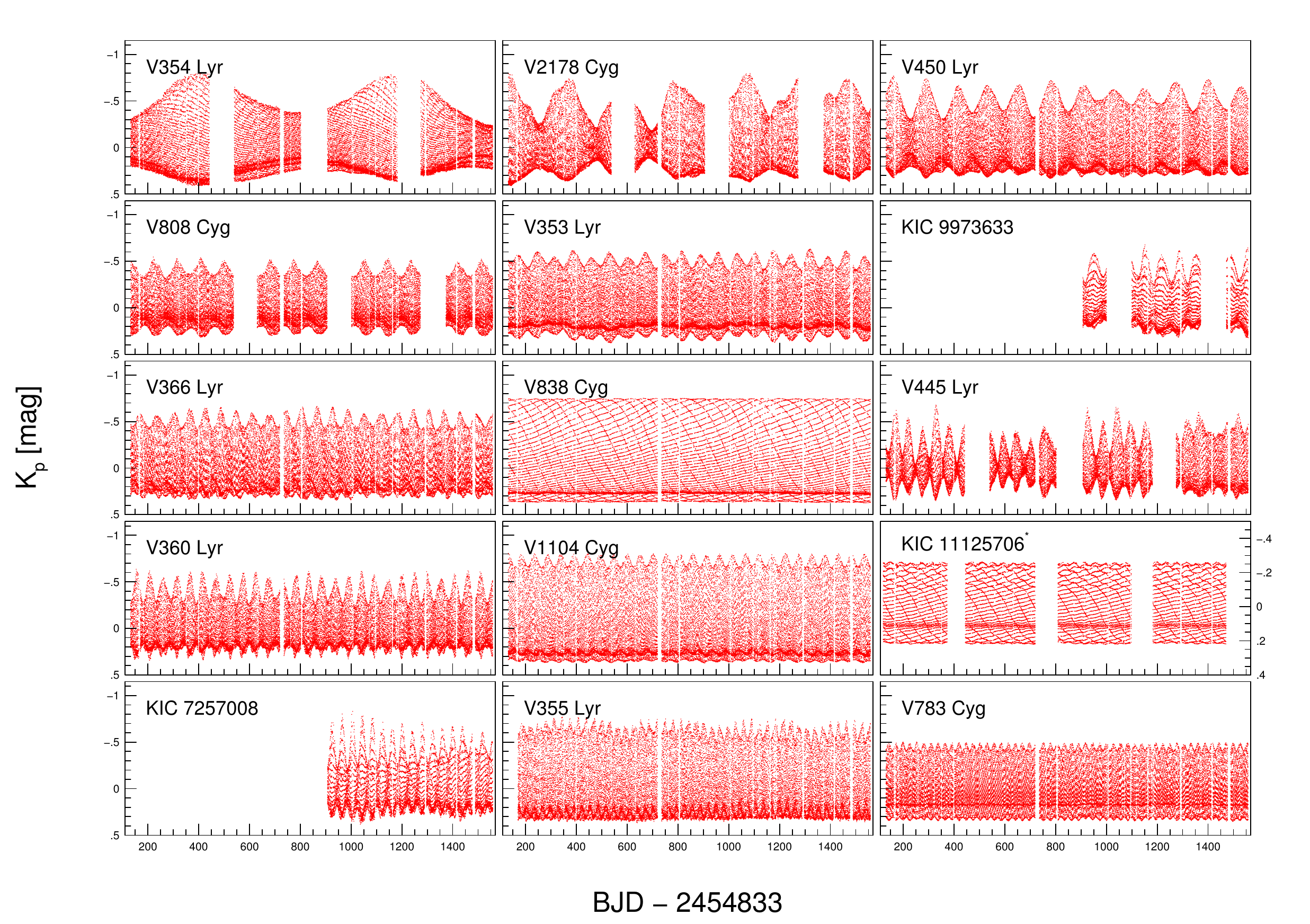}
\caption[]{
The gallery of {\it Kepler} Blazhko stars. The figure shows the complete 
light curves of fifteen stars observed with long--cadence (LC) sampling during the 
periods Q0 through Q16. The light curves are ordered by the primary Blazhko period 
from the longest (top left) to the shortest (bottom right) ones.
$^{*}$For better visibility the scale of KIC\,11125706 is 
increased by a factor of 1.5.
} \label{zoo}
\end{figure*}
By summing up the flux of all pixels in the tailor-made apertures
we obtained raw flux variation curves. 

At first sight these time series and
the {\it Kepler} flux variation curves
do not differ too much (see Fig.~\ref{llc_ny} for an example). 
Nevertheless, the difference between the flux values of the 
archived (`optimal' aperture) data
and our (`tailor-made' aperture) fluxes is considerable: about  1-5 per cents.
The exact value differs from star to star and quarter to quarter.
We note that such a comparison needs shifting the quarters 
pairwise to a common zero point.  

The main differences between the archived and our flux variation curves are that (i)
the total pulsation amplitudes $A^{\mathrm T}_i(n)$ increase for all quarters 
 ($A'^{\mathrm T}_i(n) > A^{\mathrm T}_i(n)$; Fig.~\ref{llc_2}) indicating the 
flux loss in the archived data.
Here the amplitudes $A^{\mathrm T}_i(n)$, $A'^{\mathrm T}_i(n)$ are the 
total pulsation amplitude (maximal flux $-$ minimal flux) of the $n$th pulsation 
cycle in the $i$th quarter $i,n=1,2,\dots$ for optimal and tailor-made aperture data, 
respectively. (The superscript T stands for the word `total'.)
(ii) The internal trends within quarters (see arrows in Fig.~\ref{llc_2}) 
decrease, suggesting that these trends originate from the small drift and differential velocity 
aberration of the telescope; and 
(iii) the difference of the total pulsation amplitudes
between the consecutive quarters 
$\Delta A^{\mathrm T}_{i,i+1}=\vert A^{\mathrm T}_i(l) - A^{\mathrm T}_{i+1}(1) \vert$ 
decrease: $\Delta A^{\mathrm T}_{i,i+1} > \Delta A'^{\mathrm T}_{i,i+1}$, 
(the index $l$ denotes the last pulsation cycle in the $i$th quarter). 
In an optimal case -- if the tailor-made apertures capture all the flux --
these total pulsation amplitude differences practically disappear and
only zero point shifts would remain between quarters.  

Initially we hoped that we could define tailor-made apertures
for all stars which include the total flux, i.e.
the different quarters can be joined smoothly by simple zero point shifts. 
We have found such apertures for nine Blazhko stars, only.
Six of our stars, however, show total pulsation amplitude 
differences between quarters for all possible apertures.
(For the list of the individual stars see Table~\ref{Blazhko_stars}.)
In these cases the downloaded stamps seemed to be 
too small. The right panel in Fig~\ref{Q1_map} demonstrates
such a situation well: the top pixel row and the right-most column 
contain the signal of the variable star while e.g. bottom row
does not. 

\begin{figure*}
\includegraphics[width=18cm]{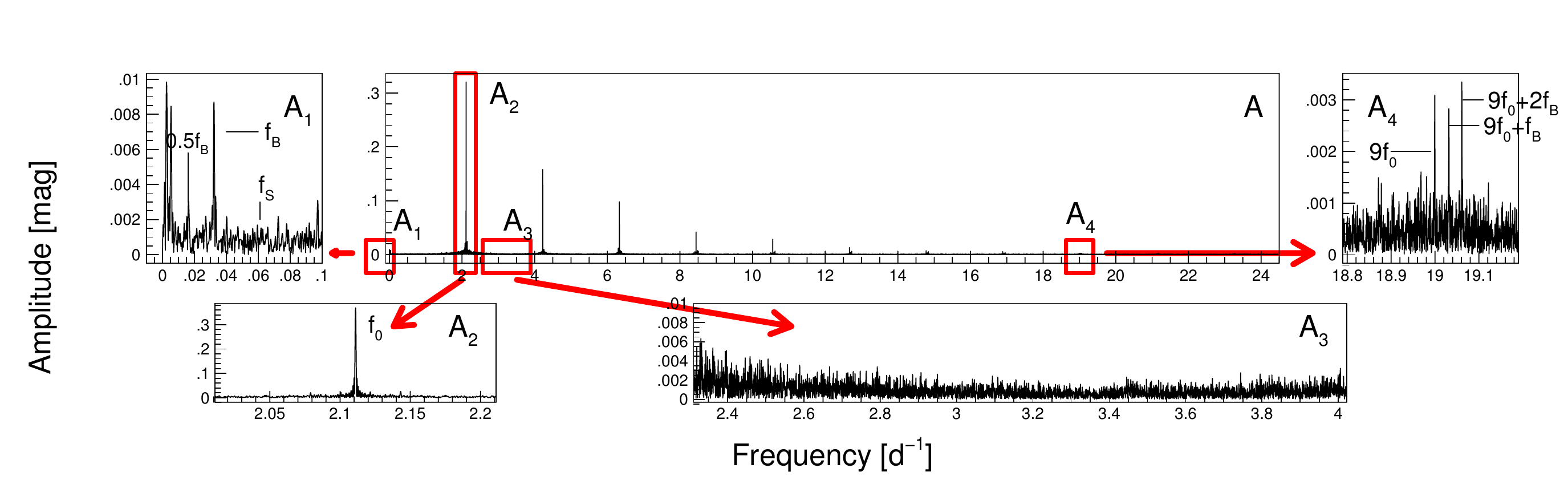}
\includegraphics[width=18cm]{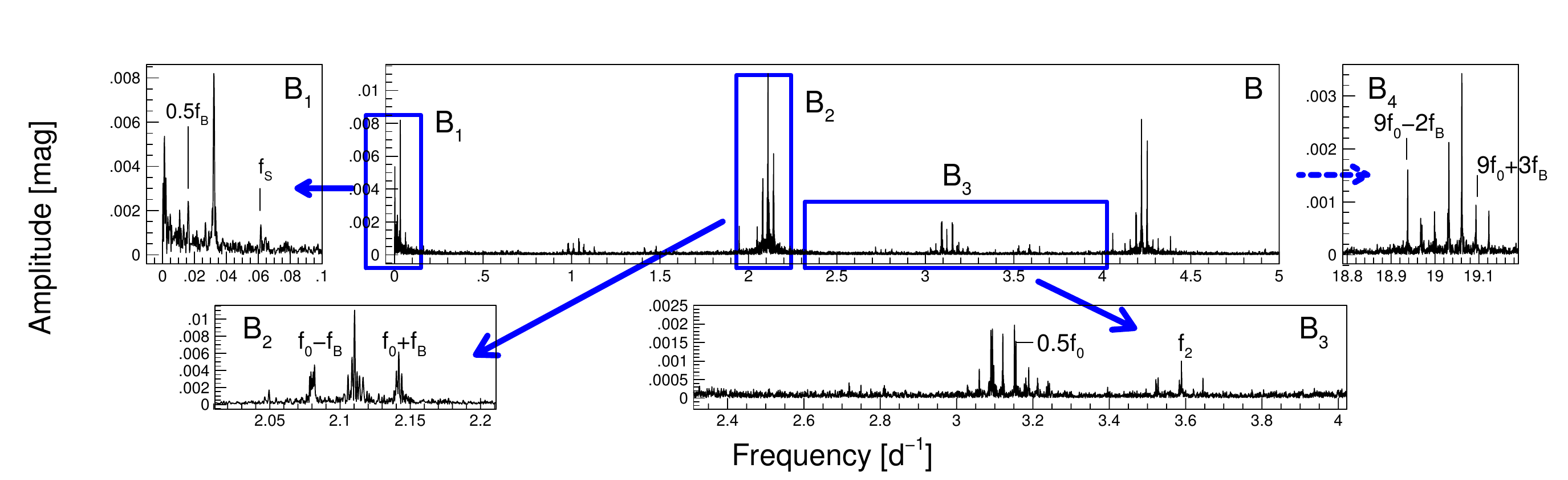}
\caption[]{
Schematic overview of the Fourier analysis of the light curves.
Top panels (signed by A): Fourier amplitude spectra, bottom panels
(signed by B) spectra after we pre-whitened the data with the main
pulsation frequency $f_0$ and its significant harmonics $kf_0$, $k=1, 2, \dots$.
Panels with indexed letters:
zooms of the the spectra, (1) low frequency range, (2) surrounding of $f_0$,
(3) additional frequencies between $f_0$ and $2f_0$, (4) high frequency range (around $9f_0$). 
Colored boxes in the middle panels show the approximate positions and 
sizes of the small (zoomed) panels. 
} \label{analysis}
\end{figure*}

How can we correct the flux variation curve of such stars?  
 Simple zero point shifts
do not result in continuous light curves, however,
we must assume that the light curve of an RR\,Lyrae 
star is continuous and smooth. To this end, 
we have to scale the flux values for properly joining the quarters.
Since at the beginning of the mission, Q4 was the most stable quarter 
(see fig.~10 in \citealt{KDCH})
we chose its fluxes as a reference for all stars. 
We defined scaling factor and zero point offset pairs 
for each quarter separately so that the transformed flux values 
of quarters can be stitched smoothly.  
These transformations are neither exact nor unique. An
additional trouble arises when one quarter of data is missing.
In those cases,
since the stamps of a star were generally fixed for the identical telescope
rolls (viz. settings for Q1=Q5=Q9,$\dots$; Q2=Q6=Q10,$\dots$ etc.), 
we used scaling factor and zero
point offset of the previous quarter at the same telescope positions:
e.g. if Q8 data are missing we use parameters of Q5 for Q9.
It must be kept in mind that this procedure may influence
the final result especially when we investigate amplitude
changes. 

The flux values of each quarter
were (1) shifted with zero point offsets and (2) multiplied by scaling factors.
Finally, (3) the eventual long time-scale trends were removed 
from  the flux variation curves by 
a trend filtering algorithm prepared for CoRoT RR\,Lyrae 
data\footnote{\url{http://www.konkoly.hu/HAG/Science/index.html}} and  (4)
fluxes were transformed into a magnitude scale, where the averaged magnitude 
of each star was set to zero. We have to emphasize that due to the logarithmic
nature of the magnitude scale all corrections and transformations should be
performed on the flux data. The measured fluxes fell between
1190 and 350\,500~e$^{-}$s$^{-1}$ which yields the estimated error  
$7.2\times 10^{-4}$ and $4.2\times 10^{-5}$\,mag 
for an individual data point, respectively.
Corrected time series data on the tailor-made apertures are available
both in flux and magnitude scales 
in electronic format\footnote{See the web page of this journal: \url{http://} \\
or \url{http://www.konkoly.hu/KIK/data.html}}.
The fluxes extracted from the tailor-made aperture
without trend filtering, applied scaling factors and offset values are 
also given in the data set (for a sample see Table~\ref{sample_data}).

\section{Analysis and Results}

\begin{figure}
\includegraphics[width=8.5cm]{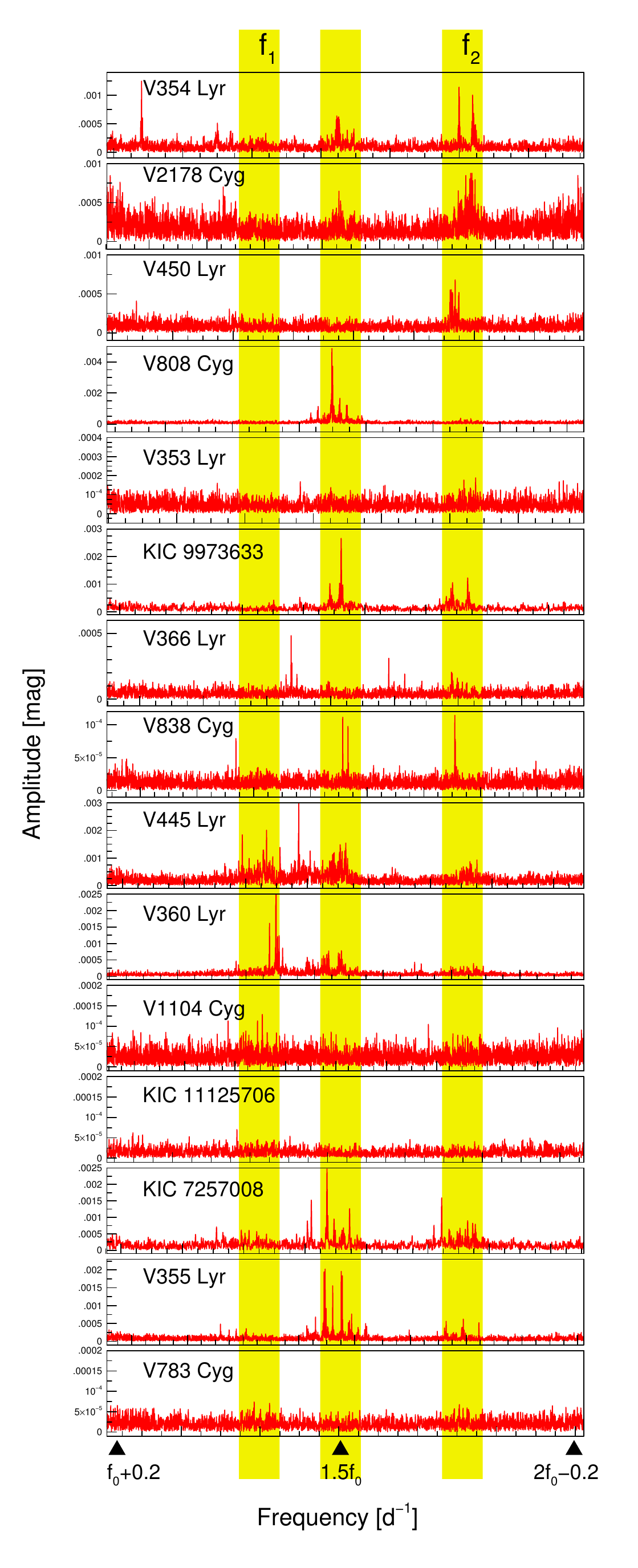}
\caption[]{
Structure of
the additional frequencies. Here we plotted the spectra between
the main frequencies and their first harmonics 
($f_0 + 0.2 < f < 2f_0 - 0.2$). All figures show residual spectra
after consecutive pre-whitening steps with the harmonics and their
numerous (3-10) significant side peaks. The yellow stripes indicate
the expected position of the radial first overtone ($f_1$), period doubling
($1.5f_0$) and radial second overtone ($f_2$) frequencies, respectively. 
} \label{fr_add}
\end{figure}

\begin{figure*}
\includegraphics[width=17cm]{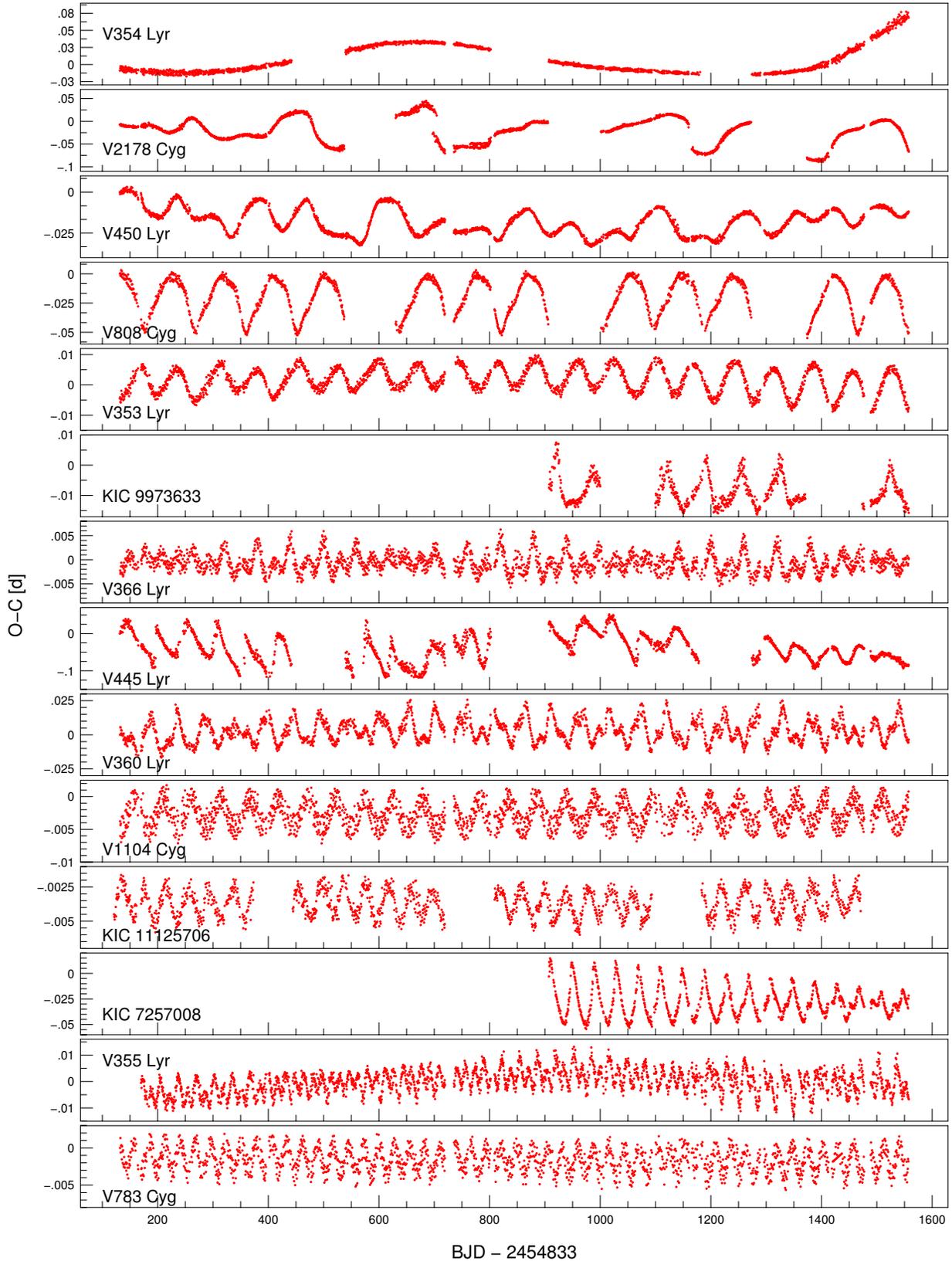}
\caption[]{The O$-$C diagrams of the analyzed Blazhko RR\,Lyrae stars.
The diagrams are constructed from interpolated times of pulsation maxima.
The basic epochs are the times of the first maximum for all stars. 
The primary AM Blazhko period decreases from top to bottom.  
The LC data of V838\,Cyg is not suitable for constructing the O$-$C diagram
(for the details see Sec.~\ref{v838cyg}), 
so it has not shown here.
} \label{o-c_all}
\end{figure*}

\begin{figure*}
\includegraphics[width=17cm]{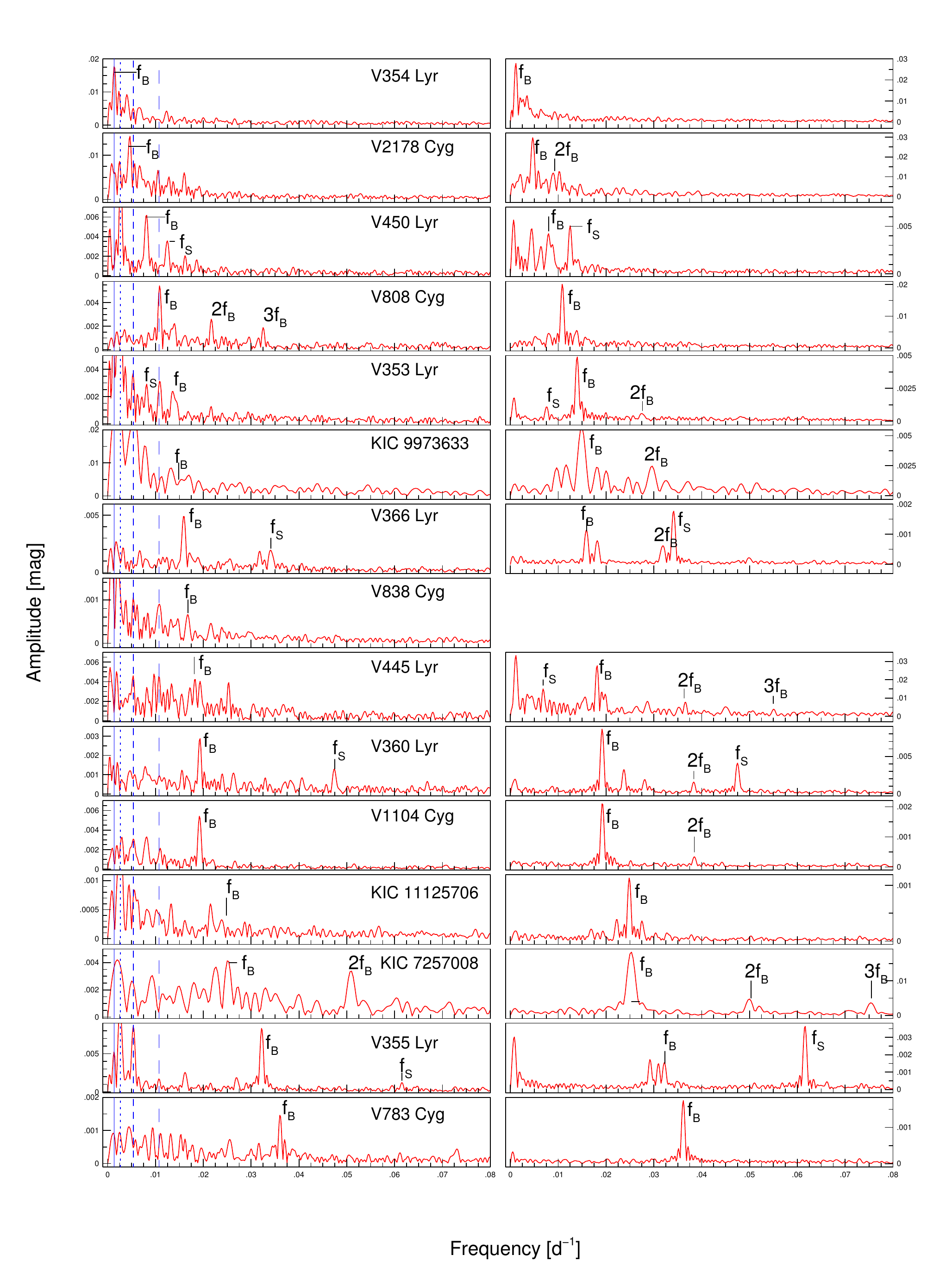}
\caption[]{Properties of amplitude and frequency modulations.
Left panels: low frequency range of the Fourier 
spectra of the light curve (Fig.~\ref{zoo}). 
Blue vertical lines show 
the location of the common instrumental frequencies: 
$f_{\mathrm K}/2$ (continuous), $f_{\mathrm K}$ (dotted), 
$2f_{\mathrm K}$ (short dashed), 
$4f_{\mathrm K}=f_{\mathrm Q}$ (long dashed), respectively.
Right panels: Fourier spectra of the O$-$C diagrams 
(Fig.~\ref{o-c_all}). 
For both cases the primary ($f_{\mathrm B}$) and secondary modulation
frequencies ($f_{\mathrm S}$) and their harmonics are marked. 
Possible linear combinations are shown in Figs.~\ref{V450_Lyr} and \ref{V366_Lyr}. 
} \label{fr_low}
\end{figure*}

\subsection{General Overview}

In the course of this study we mainly used two methods. One of them is
the Fourier analysis of the light curves that were pre-conditioned by 
the described process and shown in Fig.~\ref{zoo}. 
The second method is the analysis of O$-$C (observed minus calculated) diagrams.
In some cases other tools were used, as well. These are described later
when we discuss the relevant objects.

This paper uses the following notation conventions: numbers in the lower indices denote
the radial pulsation orders (viz. 0 = fundamental, 1 = first overtone modes, etc.).
Lower indices B and S indicate the primary (Blazhko) and secondary modulations, respectively.
Upper indices denote the detected frequencies before identification.
Throughout this paper 
the numerical values (frequencies, amplitudes, etc.) are written with the significant 
number of digits plus one digit.

\subsubsection{Fourier Analysis of the Light Curves}\label{fr_anal}

The software packages {\sc MuFrAn} \citep{Kol90} and {\sc Period04} \citep{LB05}
were used for the Fourier analysis. These program packages 
 -- together with {\sc SigSpec} \citep{Re07} --  
were tested for {\it Kepler} Blazhko stars in the 
past \citep{Benko10}. Since all of them provided similar spectra with the same
frequencies, amplitudes and phases we can use the one which 
fits the best our purposes. In this work our primary tool was {\sc MuFrAn},
but e.g. the frequency error 
or signal-to-noise ratio (S/N) were determined by {\sc Period04}.

Here we describe the general features of our Fourier analysis.
For an illustration see Fig.~\ref{analysis}.
The highest peaks in the Fourier spectra are always the main pulsation
frequencies ($f_0$) and their harmonics ($kf_0$, where $k=1, 2,\dots$) 
(see panel A in  Fig.~\ref{analysis}). 
The Nyquist frequency for the {\it Kepler} LC data is $f_{\mathrm N}=24.46$~d$^{-1}$.
Up to this limit frequency we detected 9-15 significant
harmonics depending on the pulsation frequency.

When we pre-whiten the data with these frequencies, we get Fourier spectra dominated 
by the side peaks (Fig.~\ref{analysis}B). 
The harmonics (including the main frequency) are surrounded by the 
side peaks caused by the Blazhko modulation ($kf_0\pm lf_{\rm B}$,  where $k, l=1, 2,\dots$). 
Side peaks of triplets ($l=1$) can always be seen (panel B$_2$ in Fig.~\ref{analysis}) 
and in some cases higher order multiplets ($l>1$) are also detectable (panel B$_4$). 
The higher order multiplets tend to appear around the higher order 
harmonics which indicates the frequency modulation \citep{Benko11}.

After we pre-whitened the data with a set of side frequencies 
it turned out to be evident that
the side peaks sometimes consist of double or even multiple peaks.
If we measure the spacing between these double peaks we find that
the frequency difference corresponds to the {\it Kepler} year ($P_{\rm K}=372.5$~days).
More precisely, these frequencies can be described as $kf_0\pm f'$, 
where $f'$ is one of the followings: $0.5f_{\rm K}$, $f_{\rm K}$, $2f_{\rm K}$, 
$4f_{\rm K}=f_{\rm Q}$. Here $f_{\rm K}=1/P_{\rm K}$ and 
$f_{\rm Q}=1/P_{\rm Q}$, where $P_{\rm Q}$ is the characteristic length of a quarter. 
Thus these frequencies are caused either by not properly 
eliminated problems of stitched quarters, missing quarters or 
sometimes the instrumental
amplitude variation with $P_{\rm K}$ recently discovered by \cite{Banyai13}.

In the low frequency ranges (panel B$_1$ in Fig.~\ref{analysis} 
and Fig~\ref{fr_low}) we generally find the modulation
frequencies $f_{\rm B}$, frequencies connected to the {\it Kepler} year,
and some other instrumental peaks. 
In many cases we also find the harmonics of the Blazhko frequencies
to be significant. It implies that the amplitude modulation is fairly
non-sinusoidal which is evident from the envelopes of the corresponding light curve
in Fig.~\ref{zoo}.  It is surprising that in 
many cases more than one modulation peaks can be found (see also B$_1$ in Fig.~\ref{analysis}).
These secondary modulation frequencies sometimes mimic 
to be the harmonic of the primary modulation 
frequencies but the O$-$C analysis (see Sec.~\ref{O-C_anal})
helped us to distinguish multiple modulation and non-sinusoidal
modulation. The ratios between different Blazhko frequencies
are often close to small integer numbers, which may suggests a  
resonance at work.

Although this paper focuses on the long time-scale variations, we
briefly discuss the frequency ranges between
the harmonics, since the spectra after pre-whitening with 
the modulation frequencies and the side peaks show 
specific structures (see panel B$_3$ in Fig.~\ref{analysis}).
Four stars (\object{V353 Lyr}, \object{V1104 Cyg}, KIC~11125706 and V783\,Cyg) 
do not show any significant peaks in these frequency ranges, 
but the remaining eleven stars do (Fig.~\ref{fr_add}). 

Three well separated forests of peaks can be 
identified which appear in all stars with different combinations.  
The middle ones belong to the period doubling (PD)
phenomenon \citep{Kolenberg10,Szabo10,Kollath11}.
The half-integer frequencies (HIFs: $0.5f_0, 1.5f_0,\dots$), 
their side peaks and a number of linear combination 
frequencies can be detected. If we determine the 
frequency ratio of these HIF peaks we find that their values 
frequently differ from the exact half-integer ratios. The 
explanation is a combination of mathematical, physical and sampling
effects \citep{Szabo10}. 
The frequencies located between the HIFs and $2f_0$ harmonics belong 
to the second radial overtone ($f_2$), while 
peaks between $f_0$ and HIFs are identified as 
the frequency of the first radial overtone mode ($f_1$).    
The explanation of the 
huge number of surrounding peaks around all three cases
is mathematical: the amplitude of the additional frequencies  
for both the PD effect and overtone modes  
changes in time (see e.g. \citealt{Benko10, Szabo10, Szabo14, Guggenberger12}).
Such a variable signal results in a forest 
of peaks  in the Fourier spectra as it was shown by \cite{Szabo10}.

\subsubsection{Analysis of the O$-$C Diagram}\label{O-C_anal}

The FM part of the Blazhko effect can be
separated if we study the effect in the time domain.
Since the AM and FM
definitely connected to each other, such an investigation
shows different aspects of the same phenomenon.
A practical advantage of this handling is that the time 
measurements are almost free of instrumental problems contrary to 
the brightness measurements which were discussed in Sec.~\ref{data}.

There are numerous opportunities for following frequency/period variations
from the traditional O$-$C diagram \citep{Sterken05} to the
analytic signal method \citep{Kollath02} or
we can transform it to phase variation as e.g. N13.  
We have chosen here the O$-$C diagram analysis as a simple and clear 
method. Although O$-$C diagrams were widely used for investigating 
RR\,Lyrae stars for many decades, the first diagrams that show the
period variations due to the Blazhko effect were published only
recently \citep{Chadid10, Guggenberger12},
when the continuous space-based data became available.  

As \cite{Sterken05} defined ``O$-$C stands for O[bserved] minus C[alculated]:
... it involves the evaluation and interpretation
of the discord between the measure of an observable event and its
predicted or foretold value." In our case we chose 
the time of maxima of the pulsation as an ``observable event". 
For the determination of the observed maximum times (`O' values) 
we used 7-9th order polynomial or spline fits around the maximum brightness of each
pulsation cycle. The initial epochs ($E_0$) were always the time of the first
maximum for each star. The `C' (calculated) maximum times were determined from these 
epochs and the averaged pulsation periods ($P_0$): C=$E_0 + E P_0$, where
$E=1, 2 \dots$, is the cycle number.
Gaps in the observed light curves often resulted in interpolation errors and 
consequently deviant points in the constructed O$-$C curves. We removed these points with
the {\tt time string plot} tool of Period04. The selection criterium for the wrong points was that 
they deviate from the smooth fit of the curves more than $3\sigma$, where $\sigma$
indicates the standard deviation of the fit.
The obtained curves are plotted in Fig.~\ref{o-c_all}. The accuracy of an individual O$-$C value is
about 1 minute.

The frequency content of the O$-$C diagrams were extracted again
by Fourier analysis. The corresponding spectra are shown 
in Fig.~\ref{fr_low}, where we compare them with the low frequency range of
the light curve Fourier spectra. Generally speaking the structures of these
two types of spectra are similar, but O$-$C spectra are more clear.
Here no instrumental peaks can be detected and in the case of 
multiple modulated stars the linear combination
frequencies are also more numerous and significant. 
These linear combination frequencies demonstrate
that both modulations belong to the same star
(and not to a background source) and the nonlinear coupling
between different modulations. It is noteworthy that the frequencies of
the AM and FM are always the same within their errors. 

\begin{table*}
\begin{center}
\caption{Blazhko periods and amplitudes from different methods}
\label{Blazhko_ampl}
\footnotesize{  
\begin{tabular}{*6{l@{\hspace{5pt}}}}
\tableline\tableline
Name & $P^{\mathrm{(s)}}_i$ &   $P^{\mathrm{AM}}_i$  & $A(f^{\mathrm{AM}}_i)$
& $P^{\mathrm{FM}}_i$ &  $A(f^{\mathrm{FM}}_i)$ \\  
     &   [d]           &        [d]         &        [mmag]    &        [d]      &      [min]   \\ 
\tableline
V2178\,Cyg & $207\pm15$ &   $216\pm2$ & $14\pm3.7$ & $215.9\pm0.35$ & $43.6\pm0.9$ \\ 
           & $168.8\pm1.1$ &           &       & $166.2\pm2.4$ & $18.6\pm1.2$ \\ 
V808\,Cyg  & $92.14\pm0.06$ &   $92.18\pm0.39$  & $5.4\pm0.7$  & $92.16\pm0.01$ & $30.6\pm0.1$ \\
V783\,Cyg  & $27.666\pm0.001$ &  $27.73\pm0.39$  & $1.5\pm1.2$  & $27.667\pm0.005$ & $2.6\pm0.05$ \\
V354\,Lyr  & $807\pm16$  &      &       &  $891\pm4$  & $36.9\pm0.5$ \\
V445\,Lyr  & $54.7\pm0.5$  &  $54.80\pm0.3$  & $4.3\pm1.2$  & $55.04\pm0.04$  & $38.7\pm1.5$ \\ 
           & $146.4\pm0.8$  &      &   & $147.4\pm0.7$  & $21.8\pm1.7$ \\ 
KIC~7257008& $39.51\pm0.05$ &   $39.7\pm0.4$  & $4.2\pm1.9$  & $39.72\pm0.02$ & $26.3\pm0.4$ \\
V355\,Lyr  & $31.06\pm0.1$ &   $31.04\pm0.08$  & $8.4\pm1.8$  & $30.99\pm0.02$ & $2.3\pm0.1$ \\
           & $16.243\pm0.007$ &   $16.25\pm0.1$  & $1.2\pm1.7$  & $16.229\pm0.003$ & $5.2\pm0.1$ \\
V450\,Lyr  & $123.7\pm0.4$ &   $123.0\pm1$  &  $6.5\pm1.4$  & $124.8\pm0.3$ & $5.7\pm0.3$ \\
           & $81.0\pm0.6$ &    $80.4\pm0.8$  &  $4.3\pm1.5$ & $80.1\pm0.1$  & $6.7\pm0.3$ \\
V353\,Lyr  & $71.70\pm0.04$ &   $72.1\pm1.5$  & $2.3\pm1.3$ & $71.68\pm0.02$  & $7.17\pm0.07$ \\
           & $133.1\pm0.4$  &           &       & $131.3\pm0.3$  & $1.64\pm0.08$ \\
V366\,Lyr  & $62.90\pm0.01$  &    $62.87\pm0.4$  & $5.0\pm1.4$  & $62.77\pm0.05$  & $1.66\pm0.06$ \\
           & $29.29\pm0.01$ &   $29.28\pm0.3$  & $2.1\pm1.4$  & $29.295\pm0.007$ & $2.58\pm0.06$ \\
V360\,Lyr  & $52.10\pm0.01$ &   $51.88\pm0.5$  & $2.9\pm1.2$  & $52.11\pm0.015$  & $12.9\pm0.2$ \\
           & $21.041\pm0.008$ &  $21.09\pm0.15$   & $1.3\pm1.2$  & $21.073\pm0.005$ & $6.0\pm0.2$ \\
KIC~9973633& $67.11\pm0.08$  &           &       & $67.30\pm0.07$  & $8.2\pm0.2$ \\
           & $27.13\pm0.06$  &           &       & $27.21\pm0.15$  & $1.6\pm0.4$ \\
V838\,Cyg  & $59.5\pm0.1$  &  $59.8\pm3$    & $0.6\pm2$  &        &         \\
KIC~11125706& $40.21\pm0.02$ &  &  & $40.21\pm0.01$  & $1.66\pm0.03$ \\
            &       &           &       & $58.9\pm0.1$  & $0.27\pm0.03$ \\
V1104\,Cyg & $52.00\pm0.01$  &  $52.08\pm0.2$    & $5.4\pm1.8$  & $51.99\pm0.02$  & $3.14\pm0.05$ \\
\tableline                                               
\end{tabular}
}
\tablecomments{
$P_i$ and $A(f_i)$ denote the Blazhko periods and the amplitude 
of the modulation frequencies
where $i$=B or S for primary and
secondary Blazhko periods, respectively. 
The upper indices denote the method of the calculation: (s): from 
the side peaks around harmonics; AM: direct detection
in the light curve spectra; FM: from
the spectra of the O$-$C diagrams.  
}
\end{center}                                            
\end{table*}

\subsubsection{Calculated Parameters and Accuracies}

\begin{figure}
\includegraphics[width=9cm]{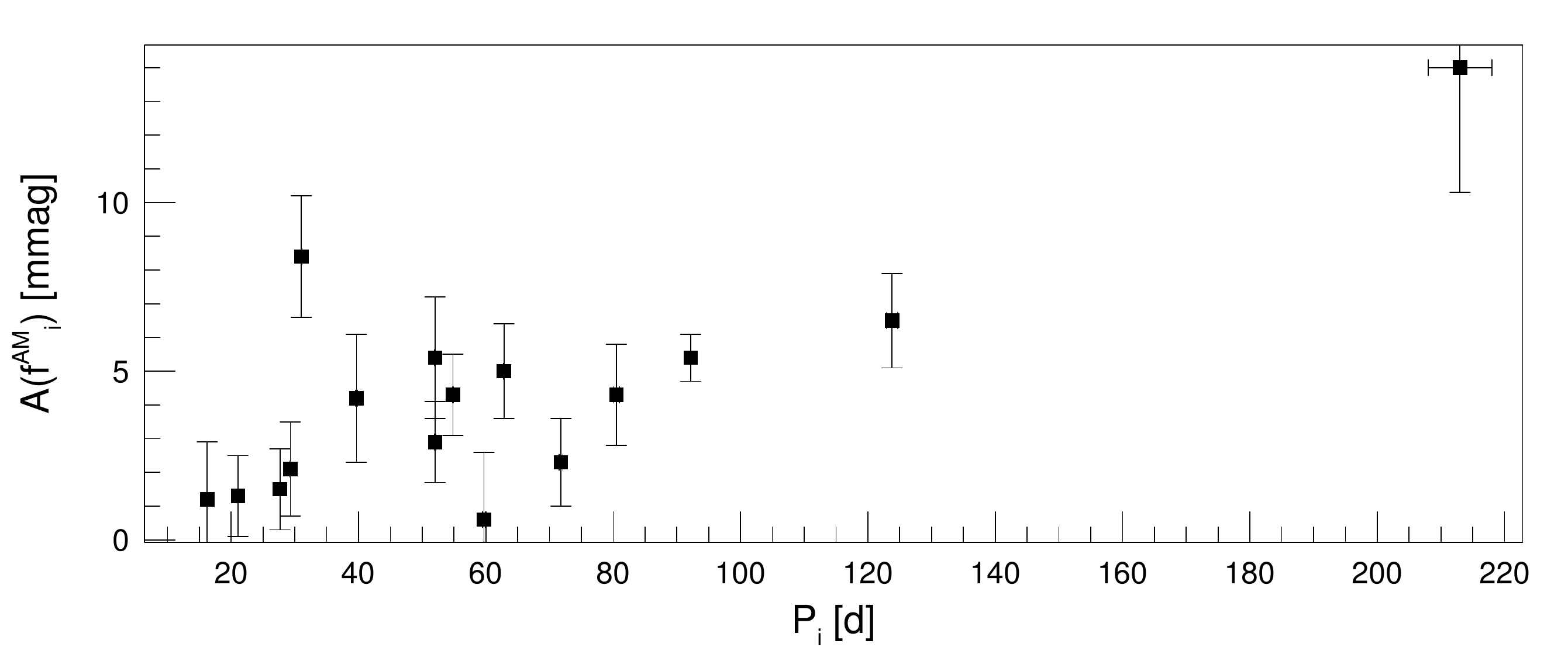}
\caption[]{Blazhko period(s) vs. amplitude 
of the AM frequency. For the plotted values see
Table~\ref{Blazhko_ampl}. 
} \label{period.vs.amplitude}
\end{figure}
The analyzed Fourier spectra have well-defined
structures and although the spectra of the light curves 
contain hundreds of significant peaks, only few of them belong to
independent frequencies. These are 
the main pulsation frequency $f_0$, the Blazhko frequencies 
($f_{\mathrm B}$ and $f_{\mathrm S}$) and the frequencies of the excited 
additional radial overtone mode(s) ($f_1$, $f_2$ and/or 
the strange mode $f_9$ which responsible for the PD effect).  
The error estimation for both the frequencies and amplitudes were 
obtained by Monte Carlo simulation
of {\sc Period04}. We note that these errors are only few percents higher
than the analytic error estimations \citep{Breger99}, because
we have almost continuous and uniformly sampled data sets.

The error estimation of the main pulsation frequency 
yields 1.1-1.8$\times 10^{-7}$~d$^{-1}$ for the Q1-Q16 data sets
while 4$\times 10^{-6}$~d$^{-1}$ for the Q10-Q16 data. These translate to
3-5$\times 10^{-8}$~d period uncertainty for the best data 
(short period and long observing span)
and $10^{-6}$~d at worst. 
The Rayleigh frequency resolution is 0.0007~d$^{-1}$ for Q1-Q16 data sets,
and 0.0015~d$^{-1}$ for Q10-Q16 data (for KIC~7257008 and KIC~9973633).   
The frequencies never change due to a modulation 
\citep{Benko11} as opposed to the amplitudes which are affected
by the FM. Consequently, our formal error estimation for the 
main pulsation amplitudes (0.3-1~mmag) are lower limits only. 

The Blazhko periods were determined by three different ways
(see Table~\ref{Blazhko_ampl}):
(i) from the averaged frequency differences of the first two triplets 
(second column);
(ii) from the Blazhko frequencies themselves (column 3)
-- if they are detectable in the spectrum of the light curve --
 and (iii) from the Fourier spectrum of the O$-$C diagrams (column 5). 
The latter two methods provide the AM and FM amplitudes
which are shown in columns 4 and 6 in Table~\ref{Blazhko_ampl}.

We call the attention of the reader to an interesting phenomenon.
In Fig.~(\ref{period.vs.amplitude}) we   
plot the Blazhko period ($P_i$) vs. amplitude
of the AM frequency ($A(f^{\mathrm AM}_i)$) diagram, where
$i$=B or S. We find a trend: the longer
Blazhko periods mean larger amplitudes and vice verse.
We can not rule out that this is a small sample effect,
 however, some arguments contradict to this scenario. The emptiness of the 
long period and small amplitude part of the diagram can be explained by
observational effect (it is difficult to distinguish between 
small-amplitude long-period stellar variations and 
the instrumental effects with similar time-scales even 
in the {\it Kepler} data), but the lack of points of the small period and
large amplitude part can not. Additionally, similar effects are 
common for (hydro)dynamical systems: e.g. weakly dissipating systems 
could be perturbed for high amplitude by long time-scale perturbing
forces only \citep{Molnar12}. The found effect will be investigated in a separate study.

Basic parameters obtained from this analysis are summarized in 
Table~\ref{Blazhko_stars}. The columns of the table show the ID numbers and
names of the stars, main pulsation periods and their Fourier amplitude
where the number of digits indicate the accuracy. Columns 5 and 6 contain the 
modulation periods averaged from the values in Table~\ref{Blazhko_ampl}. 
The last two columns of the table indicate the presence of additional frequencies, 
instrumental problems and auxiliary information about the {\it Kepler} 
observations.   

\begin{table*}
\begin{center}
\caption{Basic properties of the {\it Kepler} Blazhko stars}
\label{Blazhko_stars}
\footnotesize{
\begin{tabular}{lrlc@{\hspace{5pt}}ccrll@{}}
\tableline\tableline
KIC & GCVS  &    $P_{\mathrm 0}$ & $A(f_0)$ & $P_{\mathrm B}$&  
$P_{\rm S}$ &    &  Add. freq.\tablenotemark{a} & remarks\tablenotemark{b} \\ 
    &       &  [d] &  [mag]    & [d]  & [d]   &    &  &   \\ 
\hline
3864443 & V2178\,Cyg & $0.4869470$ & 0.3156 &  $213\pm5$              & $167.5\pm1.8$(?)     &  &  F2, (PD) & m \\
4484128 & V808\,Cyg  & $0.5478635$ & 0.2197 &  $92.16\pm0.02$         &  $\sim1000$  &  &  PD, (F2) & m \\
5559631 & V783\,Cyg  & $0.6207001$ & 0.2630 &  $27.6667\pm0.0005$  &    &           &  & scal  \\
6183128 & V354\,Lyr  & $0.5616892$ & 0.2992 &  $849\pm59$   &  (?)  &           &  F2, (PD, F') & scal, m \\
6186029 & V445\,Lyr  & $0.5130907$ & 0.2102 &  $54.83\pm0.04$  & $146.9\pm0.7$   &           &  PD, F1, F2 & m \\ 
7257008 &            & $0.511787 $ & 0.2746 &  $39.67\pm0.14$  &  $>900$  &           & PD, F2  &  Q10-\\
7505345 & V355\,Lyr  & $0.4736995$ & 0.3712 &  $31.02\pm0.05$  & $16.24\pm0.01$  &           & PD, F2 &  \\
7671081 & V450\,Lyr  & $0.5046198$ & 0.3110 &  $123.8\pm0.9$ &     $80.5\pm0.5$ &  & F2 & scal \\
9001926 & V353\,Lyr  & $0.5567997$ & 0.2842 &  $71.8\pm0.3$ &     $132.2\pm1.3$    &  &  \\
9578833 & V366\,Lyr  & $0.5270284$ & 0.2909 &  $62.84\pm0.07$ & $29.29\pm0.01$   &           & (F2) &  \\
9697825 & V360\,Lyr  & $0.5575755$ & 0.2572 &  $52.03\pm0.14$ & $21.07\pm0.03$   &           &  F2, (PD) & \\
9973633 &            & $0.510783$  & 0.2458 &  $67.2\pm 0.1$ & $27.17\pm0.06$    &           & PD, F2 & m, Q10- \\
10789273 & V838\,Cyg & $0.4802800$ & 0.3909 &  $59.7\pm0.2$ &    &           & (F2, PD) &  scal \\
11125706 &           & $0.6132200$ & 0.1806 &  $40.21\pm0.02$ & $58.9\pm0.1$  &           &  &  m, scal \\
12155928 & V1104\,Cyg& $0.4363851$ & 0.3847 &  $52.02\pm0.05$ &    &           & & scal \\
\tableline     
\end{tabular}
}
\tablecomments{
$P_0$, $P_{\mathrm B}$ and $P_{\mathrm S}$ mean the pulsation, the primary and 
secondary modulation periods, respectively. $A(f_0)$ is the Fourier amplitude of the main 
pulsation frequency.
$^{\mathrm a}$
The pattern of additional frequencies: PD means period doubling; 
F1 indicates first overtone frequency and its linear combination with
the fundamental one; F2 is as F1, but 
with the second radial overtone; F$^\prime$ indicates frequencies with unidentified modes; 
brackets indicate marginal effects.
$^{\mathrm b}$
scal=scaled, m=missing quarters, Q10- = data from Q10}
\end{center}                                            
\end{table*}

\subsection{Analysis of Individual Stars}

\subsubsection{V2178\,Cyg = KIC~3864443}\label{v2178cyg}

\begin{figure}
\includegraphics[width=9cm]{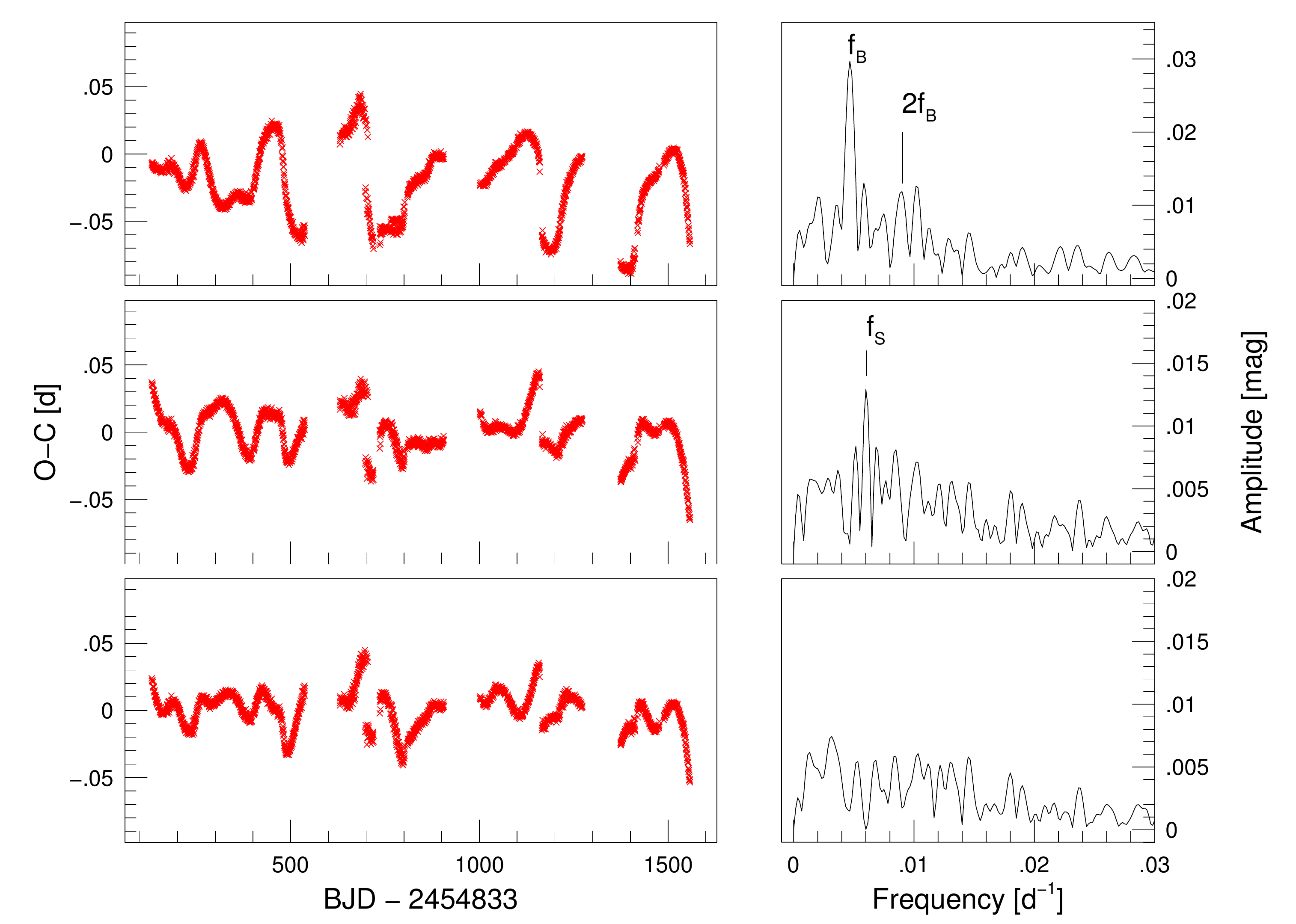}
\caption[]{
O$-$C analysis of V2178\,Cyg. Left panels  
show the O$-$C diagram (top) and its residuals after we pre-whitened 
the data with $f_{\mathrm B}$ and $2f_{\mathrm B}$ (middle)
and also with $f_{\mathrm S}$ (bottom). Right panels show
the Fourier spectra of the O$-$C data during this consecutive
pre-whitening process. 
} \label{V2178_Cyg}
\end{figure}
This star was chosen by N13 as a representative
of the long period Blazhko stars showing large AM and FM.
The envelope of the light curve in Fig.~\ref{zoo} shows 
complicated amplitude changes suggesting multiperiodicity
 and/or cycle-to-cycle variations of the Blazhko effect.
Unfortunately, the long time-scales of the variations
make quantification of this phenomenon impossible. 
We detected 11 significant harmonics of the main pulsation
frequency ($f_0$) up to the Nyquist frequency.
The triplet structures around the harmonics are highly
asymmetric: $A(kf_0-f_{\mathrm B}) \gg A(kf_0+f_{\mathrm B})$
(see also fig.~3 in N13). 
If we calculate the primary Blazhko frequency from the 
averaged spacing of the side peaks we find $0.00482$~d$^{-1}$.
This value is in good agreement with 
the highest amplitude peak in the low frequency range
($0.00462\pm 0.0001$~d$^{-1}$) so it can be identified with $f_{\mathrm B}$ 
(Fig.~\ref{fr_low}). 
Due to the missing quarters the Fourier spectrum
contains numerous instrumental frequencies such as $f_{\mathrm Q}$,
$f_{\mathrm Q}\pm f_{\mathrm B}$ and their linear combinations with 
the main frequency and its harmonics. The second largest
low frequency peak is at 0.002486~d$^{-1}$ which coincides 
with $f_{\mathrm K}=0.002685$~d$^{-1}$ within the determination error (0.0001~d$^{-1}$).

After we subtracted the largest amplitude side peaks ($kf_0-f_{\mathrm B}$)
in the spectrum of the residual, the other components of the triplets
($kf_0+f_{\mathrm B}$) and additional side peaks of 
a possible secondary modulation ($kf_0-f_{\mathrm S}$) appeared,
where $f_{\mathrm S}=0.00593$~d$^{-1}$. 
If $f_{\mathrm S}$ belongs to a secondary modulation,
the ratio of two modulation frequencies would be 2:3, however,
$f_{\mathrm S}$ can also be interpreted as a linear combination:
$f_{\mathrm S}=f_{\mathrm Q}-f_{\mathrm B}$.

When we compute the Fourier spectrum of the O$-$C diagram (Fig.~\ref{V2178_Cyg})
we find $f_{\mathrm B}=0.00463$~d$^{-1}$ and $2f_{\mathrm B}$. 
Pre-whitening with these frequencies one significant 
peak appears at 0.00602~d$^{-1}$ (S/N=24). If we remove this frequency as
well, the residual curve shows large amplitude quasi-periodic oscillations, 
but no further frequencies can be identified. Our conclusion is
that V2178\,Cyg shows a multiperiodic and/or quasi-periodic
Blazhko effect but the long periods do not allow us to draw final 
conclusion.  

Similarly to \cite{Benko10}, we found a bunch of peaks around
the frequency of the second radial overtone mode 
(see also in Fig.~\ref{fr_add}, $f_2=3.51478$~d$^{-1}$, 
$P_0/P_2=0.584$; S/N~$\approx3$). The PD
phenomenon is marginal: the highest peak around $1.5f_0$
is $f^{(1)}=3.05804$~d$^{-1}$ ($f^{(1)}/f_0=1.49$; S/N~$\approx2$).
A third additional peak condensation can be seen around
$f^{(2)}=2.656875$~d$^{-1}$ (S/N~$\approx2$). Though some Blazhko stars
(e.g. RR\,Lyr, \object{V445 Lyr}) show radial first overtone frequency 
($f_1$) around this region, we could identify this peak 
of V2178\,Cyg as a linear combination 
$f^{(2)}=3f_0-f_2$ with high certainty, because the period ratio
$P^{(2)}/P_0=0.773$ is far from the canonical value of $P_1/P_0=0.744$.
Because this period ratio increases with the increasing metallicity 
(see e.g. fig.~8. in \citealt{Chadid10})  
the measured low metallicity of  V2178\,Cyg ([Fe/H]$=-1.46$, N13) 
also supports the linear combination explanation.

\subsubsection{V808\,Cyg = KIC~4484128}\label{v808cyg}

\begin{figure}
\includegraphics[width=8.5cm]{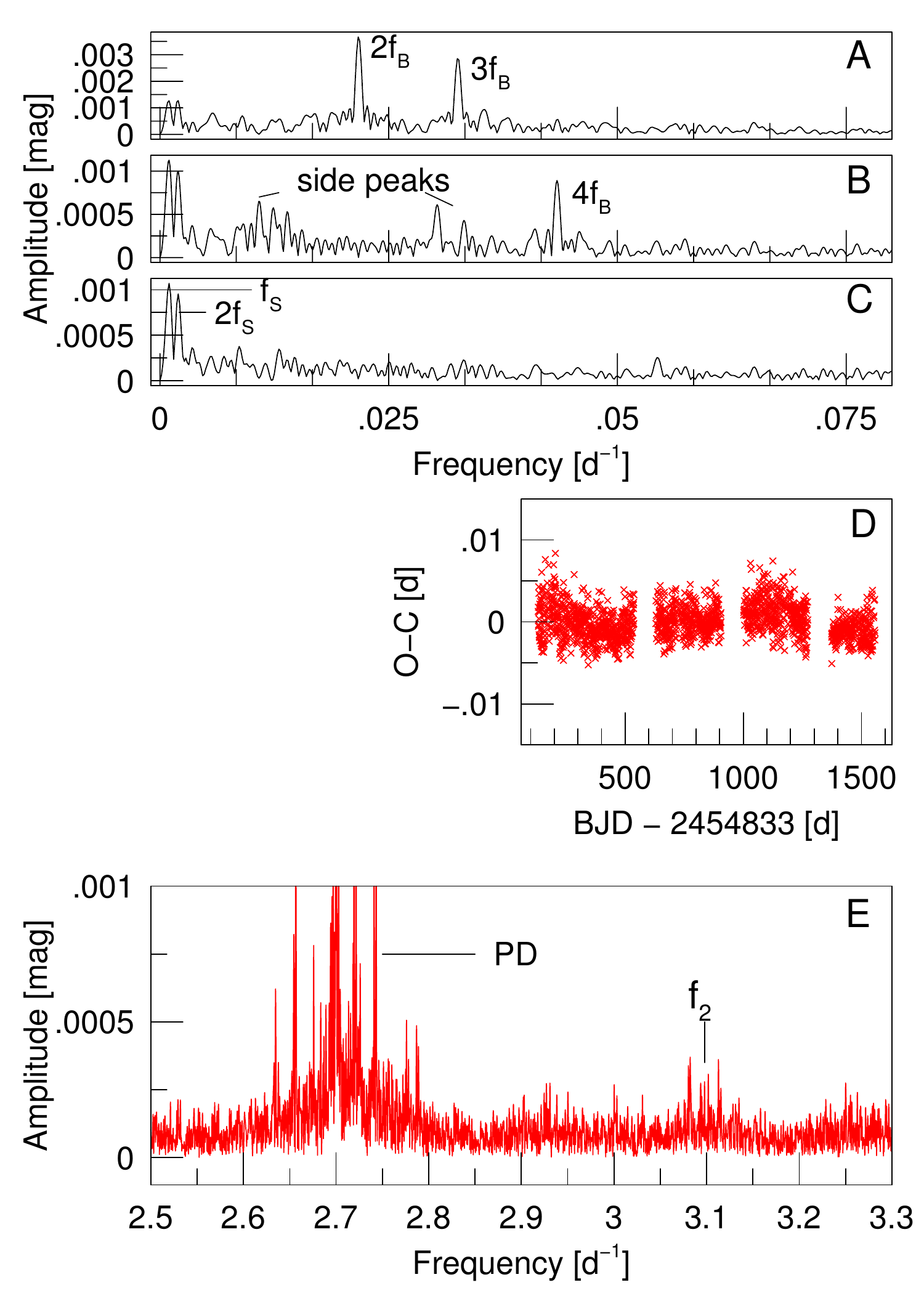}
\caption[]{
New results from V808\,Cyg.  A-C panels: 
The pre-whitening process of the O$-$C diagram shown in 
Fig.~\ref{o-c_all}. The harmonics
of the Blazhko frequency $f_{\mathrm B}$ are significant up 
to the 4th order. A long time-scale variation is evident 
because of the presence of 
low frequency peaks ($f_{\mathrm S}$, $2f_{\mathrm S}$) and
side frequencies around the harmonics of $f_{\mathrm B}$.
Panel D: residual O$-$C curve containing only 
$f_{\mathrm S}$ and $2f_{\mathrm S}$. Panel E: the second
radial overtone frequency ($f_2$) can be detected from the Q1-Q16 data.
} \label{V808_Cyg}
\end{figure}
The light curve of this star in Fig.~\ref{zoo} shows
two important features. First, the
envelope shape suggests a highly non-sinusoidal AM.
Second, the length of the Blazhko cycle is close to
the length of the observing quarters.
As a consequence of the first fact, we can detect two
significant harmonics $2f_{\mathrm B}$ and $3f_{\mathrm B}$ 
of the Blazhko frequency $f_{\mathrm B}=0.01085$~d$^{-1}$ 
(Fig.~\ref{fr_low}), and
multiplet side peaks ($kf_0\pm lf_{\mathrm B}$, where $l>1$)
are detectable, as well. A slight cycle-to-cycle amplitude change 
might be present but 
the quarter-long Blazhko period and gaps together make
such an effect barely detectable.

The O$-$C diagram of \object{V808 Cyg} can be fitted well
with the Blazhko period and its three harmonics.
After we subtracted this four-frequency fit from the O$-$C data,
a definite structure can be detected in the residual spectrum 
(panel B in Fig.~\ref{V808_Cyg}). 
At the position of $f_{\mathrm B}$ and $3f_{\mathrm B}$ 
side peaks appear. These peaks define a secondary modulation 
with the frequency of $f_{\mathrm S}=0.0010$~d$^{-1}$.
The pre-whitened spectrum indeed shows two peaks at $f_{\mathrm S}$ and $2f_{\mathrm S}$ 
(panel C).
However, the possible period $P_{\mathrm S}\sim 1000$~d is commensurable with 
the length of the total observational time, so this O$-$C variation in panel D could
be secular, as well.

V808\,Cyg shows the strongest known PD effect that is why
data taken during the first two quarters were investigated in detail 
by \cite{Szabo10}. Using the time series up to Q16 
this main finding remains unchanged. The highest amplitude
HIF is at $1.5f_0$, namely $f^{(1)}=2.69770$~d$^{-1}$ 
($f^{(1)}/f_0=1.48$; S/N~$\approx 30$). 
After applying a few-step pre-whitening process -- when we subtract the 
main pulsation frequency, its harmonics and some (6-10) significant
multiplets around the harmonics -- we found that the second
radial overtone mode (or a non-radial mode at the location of the
radial one) $f_2=3.09774$~d$^{-1}$ ($P_2/P_0$=0.589) is also excited
(panel E in Fig.~\ref{V808_Cyg}). 
The amplitude of this frequency is much lower than the amplitudes 
of the PD frequencies, which explains why previous investigations have 
not discovered this mode.

\subsubsection{V783\,Cyg = KIC~5559631}\label{v783cyg}

\begin{figure*}
\includegraphics[width=17cm]{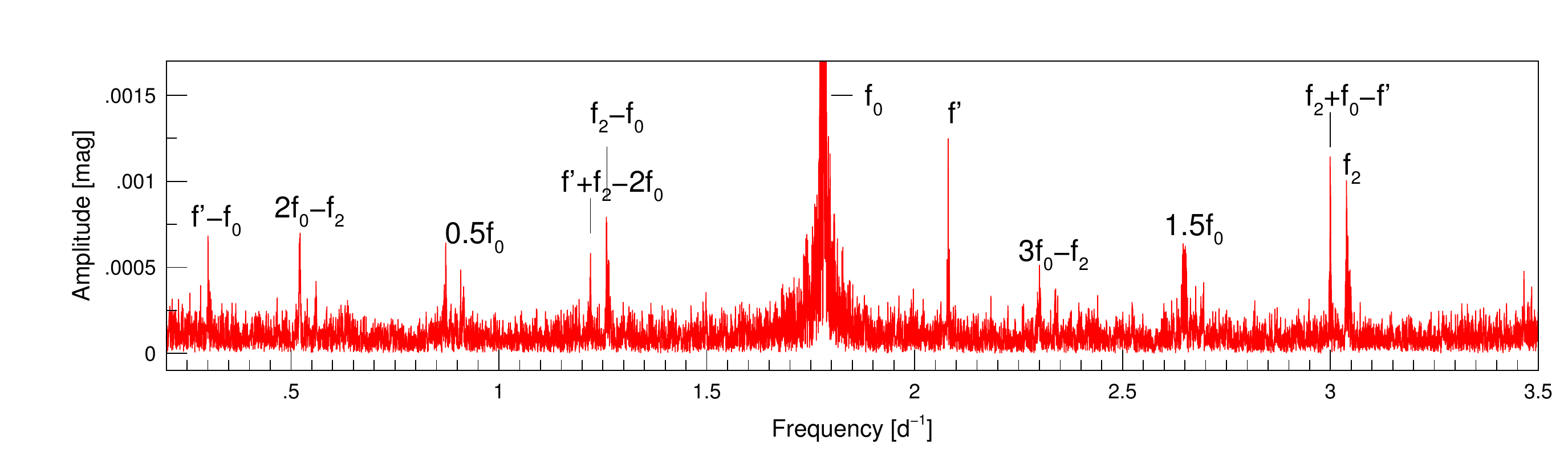}
\caption[]{
Additional peaks in the lower frequency range of
the pre-whitened Fourier spectrum of  V354\,Lyr light curve. 
} \label{V354_Lyr}
\end{figure*}
The Blazhko effect of V783\,Cyg seems to be  
simple: a sinusoidal AM and FM
visible both in the light curve (Fig.~\ref{zoo}) and O$-$C diagram 
(Fig.~\ref{o-c_all}). By investigating these curves more carefully
 we can detect small differences between consecutive cycles.   

When we pre-whiten the light curve data with the main pulsation 
frequency and its 15 significant harmonics we find nice
symmetric triplet structures in the spectrum around the harmonics. 
The star has the shortest Blazhko 
cycle in our sample ($P_{\mathrm B}=27.67$~d).
The spectrum also contains the modulation frequency itself: 
$f_{\mathrm B}=0.036058$d$^{-1}$. After 
subtracting the triplets in the residual spectrum,
multiplet side frequencies appear. By carrying out a similar
few-step pre-whitening process as in Sec.~\ref{v808cyg} we can eliminate all side peaks and no
additional peaks emerge between the harmonics. 

Fourier analysis of the O$-$C diagram provides us $f_{\mathrm B}$
again (Fig.~\ref{fr_low}). When we subtract a fit with  $f_{\mathrm B}$, 
the residual O$-$C diagram shows a parabolic shape indicating 
a period change. Fitting a quadratic function in the form 
\[
\mathrm{O}-\mathrm{C}=\frac{1}{2} \frac{dP_0}{dt} \bar{P_0} E^2 
\]
\citep{Sterken05}, where $\bar{P_0}$ means the averaged period,
$E$ the cycle number from the initial epoch, we find 
a period increase: $dP_0/dt=1.02\times 10^{-9}\pm1.7\times 10^{-10}$~dd$^{-1}$.
That is $0.12\pm0.02$~dMy$^{-1}$ which agrees well with the value 
of \cite{Cross91} $0.088\pm0.023$~dMy$^{-1}$, who determined it on the 
basis of photometric observations between 1933 and 1990.

The Fourier analysis of the light curve and O$-$C diagram 
are not sensitive for the mentioned slight cycle-to-cycle variation
of the Blazhko effect. 
The short Blazhko period and uninterrupted {\it Kepler} data 
make V783\,Cyg a good candidate for a dynamical analysis.
This study will be discussed in a separate paper by
Plachy et al. (in prep.). The preliminary results 
suggest that the cycle-to-cycle  
variation of V783\,Cyg light curve and O$-$C 
diagram have chaotic nature.

\subsubsection{V354\,Lyr = KIC~6183128}\label{v354lyr}

The star has the longest Blazhko period in
the {\it Kepler} sample. 
We find $P_{\mathrm B}=807$~d
($f_{\mathrm B}=0.00124$~d$^{-1}$) calculating from 
the triplet spacing in the spectrum of the light curve. 
At the same time the highest peak in the low frequency range is
at $f_{\mathrm B}=0.00134$~d$^{-1}$ which yields
$P_{\mathrm B}=748$~d. The problem is that 
the Blazhko period is inseparably close to the instrumental
period $2P_{\mathrm K}=745$~d. 

The observed two Blazhko cycles in Fig.~\ref{zoo} 
look different. The ascending branch of the first
cycle is steeper than that of the second one while  
the descending branch of the second cycle is the steeper.
The different shapes of the two cycles 
in the O$-$C diagram (Fig.~\ref{o-c_all}) strengthen our suspicion:
\object{V354 Lyr} shows multiperiodicity or cycle-to-cycle variations in the
Blazhko effect. The long Blazhko period
prevents us from quantifying this feature.

As it was already found by \cite{Benko10}, the Fourier
spectrum of V354~Lyr contains significant additional
peaks between harmonics. The following frequencies 
were reported between $f_0$ and $2f_0$ (with the notations of the referred paper): 
$f'=2.0810$, $f''=2.4407$, $f'''=2.6513$ and $f_2=3.03935$~d$^{-1}$. 
After we removed the main pulsation frequency, its significant 
harmonics, and the highest amplitude side peaks around the harmonics,
our spectrum (Fig.~\ref{V354_Lyr}) also contains numerous additional
frequencies. The highest amplitude peak from the half-integer frequency is 
located now at $f^{(1)}=2.648387=1.5f_0$ ($f^{(1)}/f_0$=1.49).
Identifying the frequency of the radial second overtone mode is problematic,
because there is a double peak in that position: 
$f^{(1)}_2=3.038671$~d$^{-1}$
($P^{(1)}_2/P_0=0.586$) and $f^{(2)}_2=2.999333$~d$^{-1}$ 
($P^{(1)}_2/P_0=0.593$). The spacing between these two frequencies
is 0.300d$^{-1}$. Similar double peak structures can be seen 
for many other frequencies. 

The highest amplitude additional frequency $f'=2.080672$~d$^{-1}$ 
is mysterious. We have not found any frequencies in such a position for
other Blazhko stars (see also Fig.~\ref{fr_add}).
Neither known instrumental frequencies nor linear combinations
of instrumental and real frequencies give peak at this position.
The spectral window function has a comb like structure \citep{KIH},
but the peaks are far from $f'$ and their amplitude 
are very low (in a normalized scale $\sim 0.004$).
So we can rule out $f'$ being an instrumental frequency.  

We checked whether this frequency comes
from this star or not. Using the flux data we searched for
its linear combination frequencies with the main pulsation one.
Both the $f'+f_0=3.8613$~d$^{-1}$ and $f'-f_0=0.3003$~d$^{-1}$  are indeed
detectable. So we could rule out a background star as the 
source of this frequency. The period ratio is $P'/P_0=0.855$.
\cite{Benko14} raised an explanation that this frequency 
would be a linear combination of the main pulsation
$f_0$ and radial first overtone mode frequencies $f_1$, namely
$f'=(f_0+f_1)/2$. The problem of this interpretation 
is that the spectrum shows significant peaks neither at $f_0+f_1$ nor at $f_1$.

As we have seen the spacing between $f^{(1)}_2$ and $f^{(2)}_2$ can
be given as $f' - f_0$, so any of them can be calculated as a linear
combination of $f_2$, $f_0$ and $f'$. 
In Fig.~\ref{V354_Lyr} we identified $f_2=f^{(1)}_2$, therefore
$f^{(2)}_2=f_2+f_0 - f'$. Alternatively, if we identify $f_2=f^{(2)}_2$,
then $f^{(1)}_2=f_2+f' - f_0$.  
We can find several other peaks where we have alternate identifications
depending on how we choose $f_2$.

The last frequency listed by \cite{Benko10} is $f''=2.4407$~d$^{-1}$.
Although this frequency is not significant in the spectrum
calculated from Q1-Q16 data, but the half of it (1.220~d$^{-1}$) can be detected.
The later frequency can easily be produced by $f_2-f_0$ if we identify
$f_2=f^{(2)}_2$.  As we have shown in \cite{Benko14} frequency combination
$2(f_2-f_0)$ could explain numerous previously unidentified peak 
in spectra of some {\it CoRoT} and {\it Kepler} Blazhko stars.
So it is not surprising that we temporarily detect this 
combination frequency $2(f_2-f_0)=2.4407$~d$^{-1}$ in V354~Lyr, as well.
As we mentioned in Sec.~\ref{fr_anal} all additional
frequency amplitudes (HIFs, overtones) change in time,  
these combination frequencies also seem to show similar time dependency.

\subsubsection{V445\,Lyr = KIC~6186029}\label{v445lyr}

The light curve of the star shows  strong
and complicated amplitude changes (Fig.~\ref{zoo}). 
It was the subject of a detailed study of \cite{Guggenberger12}.
The paper used the data available in that time    
(Q1-Q7), but its main statements are not changed in the 
light of the more extended Q1-Q16 data set.
The heavily varying parameters such as the different periods,
amplitudes and phases result in a slightly different
averaged values for these parameters compared to
ones given by \cite{Guggenberger12}. We confirm
the existence of two modulation frequencies ($f_{\mathrm B}$,
$f_{\mathrm S}$) and four additional frequency patterns,
namely $f_2$, $f_1$, PD and $f_{\mathrm N}=2.763622$~d$^{-1}$.
In the latter case we noted a possible interpretation with
the linear combination of $f_{\mathrm N}=2(f_2-f_0)$ in \cite{Benko14}.

\subsubsection{KIC~7257008}\label{kic7257008}

The variable nature of the star was discovered by the
ASAS survey (\citealt{Poj97, Poj02}; Szab\'o et al. in prep.). 
Its {\it Kepler} data were investigated for 
the first time by N13. The envelope of the
light curve in Fig.~\ref{zoo} suggests multiple modulation
behavior. We determined the Blazhko frequency both 
from the side peak patterns and from the low frequency range of
the Fourier spectrum of the light curve: 
$f_{\mathrm B}$=0.02528~d$^{-1}$. The harmonic $2f_{\mathrm B}$
is also significant as a consequence of the non-sinusoidal nature 
of the AM. 

The FM is even more
non-sinusoidal: the Fourier spectrum of the O$-$C diagram (in Fig.~\ref{fr_low})
contains five significant harmonics of the Blazhko frequency.
A small peak at the sub-harmonic 
$f_{\mathrm B}/2=0.01234$~d$^{-1}$ (S/N=3.6) is also present.
After pre-whitening the Fourier spectra of the upper envelope (maxima) curve
or O$-$C diagram, double side peaks remain at the
location of the Blazhko frequency and its harmonics.
The spacing between side peaks is very narrow ($\sim 0.001$~d$^{-1}$) 
which implies a secondary Blazhko period longer than the length of the 
data set (Q10-Q16). The amplitude and the phase of the harmonics of
$f_{\mathrm B}$ have changed during the observing term producing
varying envelope and O$-$C curves. These changes have been verified
by using {\sc Period04} amplitude and phase variation tool.
\cite{Molnar14} have found that the star shows PD effect
($1.5f_0=2.871047$~d$^{-1}$) and contains the second overtone
pulsation ($f_2=3.329353$~d$^{-1}$) as well.

\subsubsection{V355\,Lyr = KIC~7505345}\label{v355lyr}

\begin{figure}
\includegraphics[width=4cm,angle=-90,trim=100 30 100 100]{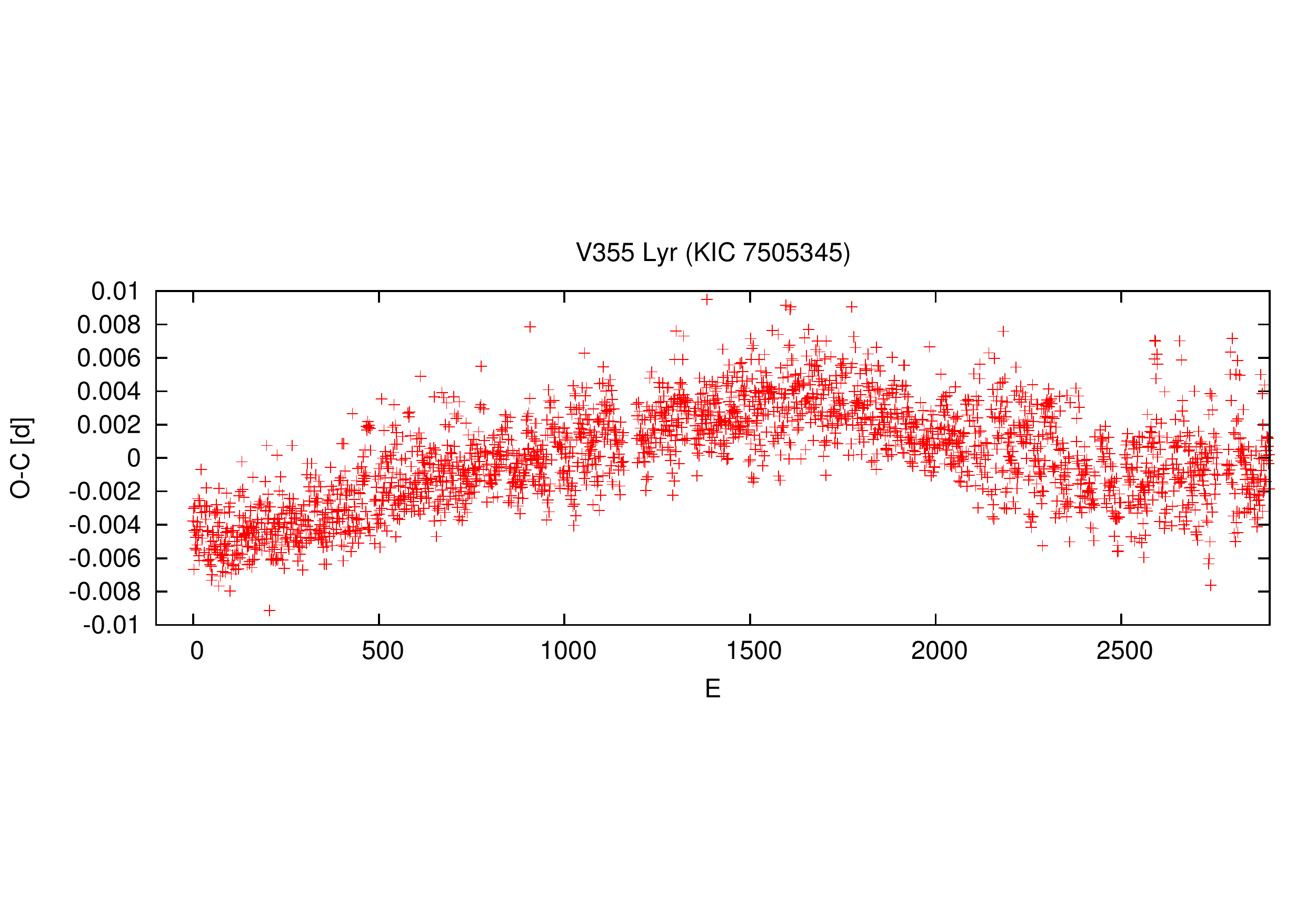}
\caption[]{
Residual O$-$C diagram of V355\,Lyr after we pre-whitened
the data with the frequencies $kf_{\mathrm B}$, ($k=1,2,3,5$),
and side peaks $f_{\mathrm B}\pm f_{\mathrm L}$.
} \label{7505345_o-c}
\end{figure}
The light curve of \object{V355 Lyr} in Fig.~\ref{zoo} suggests at 
least two modulation periods. The longer period amplitude
change shows about four cycles during the four years observing
span. It raises the possibility of an instrumental effect showing the 
Kepler year (372.5~d) \citep{Banyai13}. Indeed, we found two strong peaks
in the low frequency range of the spectrum which can be
identified as $f_{\mathrm K}=0.00266$ and $2f_{\mathrm K}=0.00533$~d$^{-1}$.
But other aspects contradict such an explanation.
The Blazhko frequency turns up from the triplet 
structures and it is detectable directly as $f_{\mathrm 
B}=0.0322\pm0.005$~d$^{-1}$.
There is a detectable peak at $f=0.06154$~d$^{-1}$, (S/N=4.5) which 
 can not be the harmonic $2f_{\mathrm B}$, because its difference
from the exact harmonic (0.00147~d$^{-1}$) is twice of the 
Rayleigh frequency resolution ($\approx0.0007$~d$^{-1}$).
So we may identify it as a possible secondary modulation 
frequency ($f=f_{\mathrm S}$). If it is true, the two modulation
frequencies are in a nearly 1:2 ratio, causing the 
observed beating phenomenon in the envelope curve.

At the same time the sub-harmonic $f_{\mathrm B}/2$ is also significant.
A similar situation has been found for the first time by \cite{Sodor11} 
in the case of the multiperodic Blazhko star \object{CZ Lac}. 
\cite{Jurcsik12} also detected the sub-harmonic of the Blazhko 
frequency for \object{RZ Lyr}. As \cite{Sodor11} discussed, we can also identify 
$f_{\mathrm B}/2$ as the primary modulation frequency. In that case 
instead of a sub-harmonic we would have a harmonic with much higher amplitude than
the main modulation frequency. Moreover, the modulation curve would have a
rather unusual shape. For these reasons we prefer the sub-harmonic
identification. 

As opposed to the above mentioned amplitude relations, the Fourier amplitude 
$A(f_{\mathrm S})$ is higher than $A(f_{\mathrm B})$ in the spectrum of 
the O$-$C curve.
In other words, in FM $f_{\mathrm S}$ dominates, while in AM $f_{\mathrm B}$
is the dominant.
Such an effect have never been detected before.
Linear combination peaks at  
$f_{\mathrm S}\pm f_{\mathrm B}$ 
are also detectable. Additionally, the sub-harmonic of $f_{\mathrm B}$ can not be seen, but
the sub-harmonic of $f_{\mathrm S}=0.03083$~d$^{-1}$ can be detected.
One other significant peak is at $f^{(1)}=0.16316$~d$^{-1}$ (S/N=5.3).
This frequency is located close to $5f_{\mathrm B}=0.16150$~d$^{-1}$ but
the identification $f^{(1)}=5f_{\mathrm B}$ is ambiguous, because no other harmonics 
are detectable. The spectrum of the light curve also shows
a marginal (S/N=2.4) peak at this position. 
We suspect that it might be a third modulation frequency.  
Pre-whitening the O$-$C curve with all the above mentioned 
frequencies we obtain the residual curve 
in Fig.~\ref{7505345_o-c}. This residual
shows a sudden period change at about $E\approx1636$ (= BJD$\approx2455778$) and a
less pronounced one around $E\approx2386$ (= BJD$\approx2456133$). Nothing particular can be seen
in the light curves around these dates.

Higher frequency range of the time series data is
dominated by the main pulsation frequency, its harmonics, and
their strong multiplet surroundings.
Beyond the clear PD effect ($1.5f_0=3.155484$~d$^{-1}$; 
$P/P_0=1.495$) discussed by \cite{Szabo10},
the Fourier spectrum of the star shows evidences for
the second radial overtone pulsation (see in Fig.~\ref{fr_add}). 
This feature was undetected
by previous {\it Kepler} studies. The frequency $f_2=3.589528$~d$^{-1}$
($P_2/P_0=0.588$) is surrounded by well separated side peaks.

\subsubsection{V450\,Lyr = KIC~7671081}\label{v450lyr}

\begin{figure}
\includegraphics[width=8.5cm]{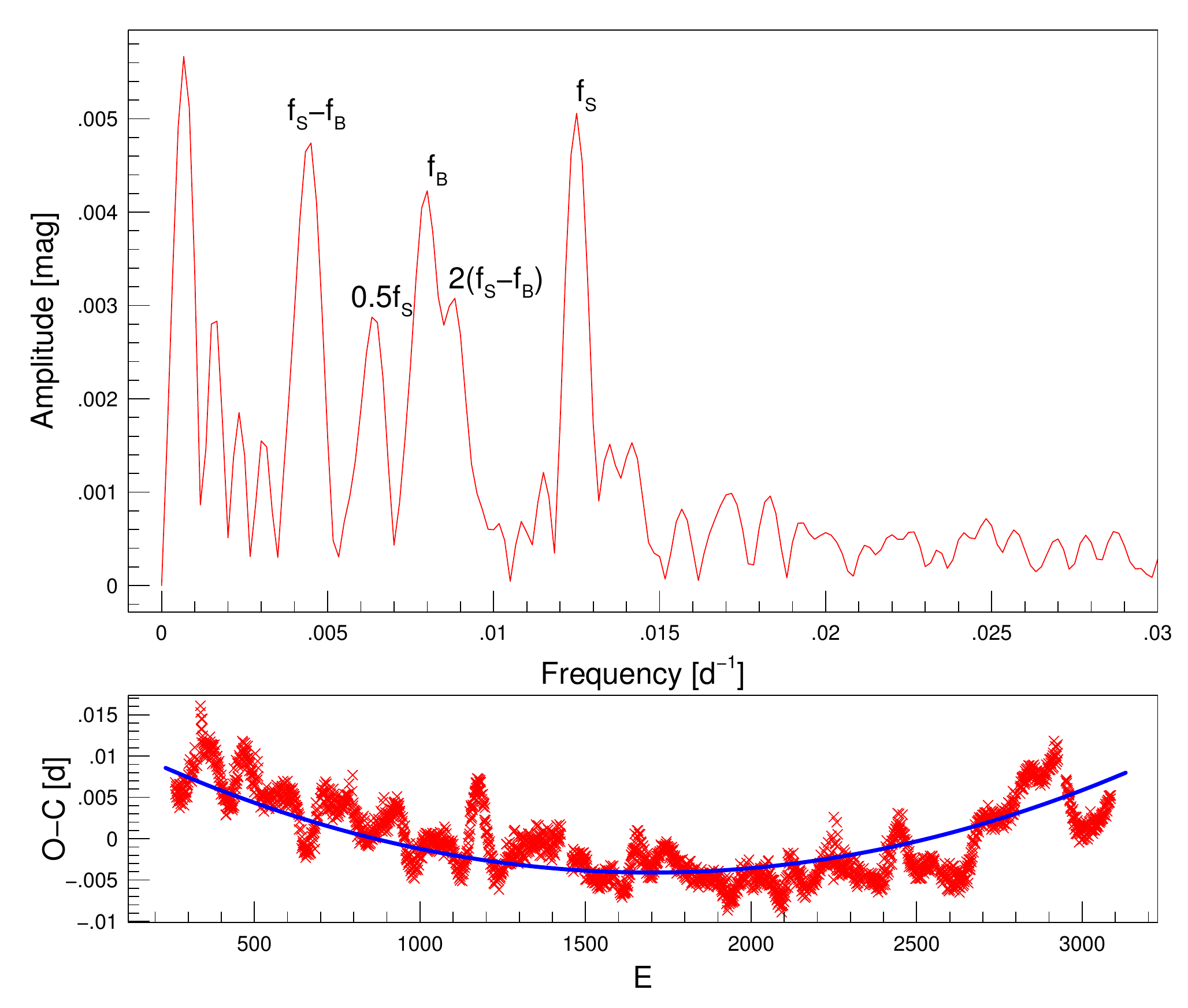}
\caption[]{
Top: Zoom from the Fourier spectrum of the O$-$C diagram of V450\,Lyr.
Bottom: Residual O$-$C diagram after we pre-whitened
the data with the above signed five frequencies. The best fitted parabola
(continuous line) suggests a very fast period increase.
} \label{V450_Lyr}
\end{figure}
The shape of the maxima curve of \object{V450 Lyr} suggests a strong beating 
phenomenon between two modulation periods, however, 
similar features have also been seen in other stars 
(e.g. V355\,Lyr and KIC~7257008) which proved to be
instrumental effects. Accordingly, we carefully compared the 
frequencies determined in the spectra of the light curve and the O$-$C diagram.
In the case of the light curve
the largest low frequency peak $f_{\mathrm K}$ 
belongs to the Kepler year. 
The next one is the modulation frequency $f_{\mathrm B}=0.00813$~d$^{-1}$
($A(f_{\mathrm B})$=6~mmag). 
Its harmonic $2f_{\mathrm B}$ is also detectable. The third largest 
amplitude peak is at the frequency $f_{\mathrm S}=0.01243$~d$^{-1}$.
Low significance peaks are seen at the $f_{\mathrm B}+f_{\mathrm S}$ 
and $f_{\mathrm S}/2$ 
($A(f_{\mathrm B}+f_{\mathrm S})\approx A(f_{\mathrm S}/2)\approx2$~mmag).  

A similar analysis on the spectrum of O$-$C diagram (see top panel in 
Fig.~\ref{V450_Lyr})
results in two independent frequencies $f_{\mathrm B}$, $f_{\mathrm S}$ and
some combinations of them ($f_{\mathrm S}-f_{\mathrm B}$, 
$f_{\mathrm S}-2f_{\mathrm B}$, $f_{\mathrm S}/2$).
We can state now that $f_{\mathrm S}$ or $f_{\mathrm S}/2$ is a real 
secondary modulation frequency (See the discussion about 
the interpretation of the residual sub-harmonic
in Sec.~\ref{v355lyr}).  
When we subtract a fit with all the mentioned frequencies, 
the residual O$-$C diagram (bottom panel in Fig.~\ref{V450_Lyr}) shows a combination of
remained quasi-periodic signal and a parabolic shape indicating 
a strong period change. Fitting a quadratic function we found
a fast period increase: $dP_0/dt=2.4\times 10^{-8}$~dd$^{-1}$, which can
not be caused by stellar evolution.
This phenomenon can be explained e.g. as a sign of a third longer period
modulation or as a random walk caused by a quasi-periodic/chaotic
modulation.

By investigating the fine structure of the Fourier spectrum 
between harmonics of the main pulsation frequency
\cite{Molnar14} recognized that V450\,Lyr pulsates in the
second radial overtone mode, as well (see also Fig.~\ref{fr_add}). 
If we identify the highest peak in this region at 3.33670~d$^{-1}$ 
with $f_2$, the period ratio is $P_2/P_0=0.594$, corresponding
to canonical second overtone period ratio. 

\subsubsection{V353\,Lyr = KIC~9001926}\label{v353lyr}

The AM of V353\,Lyr displays alternating 
higher and lower amplitude Blazhko cycles (Fig.~\ref{zoo}).
The phenomenon reminds us the PD effect where the amplitudes of the
consecutive pulsation cycles alternate. The effect suggests
two modulation frequencies in a nearly 1:2 ratio.
The low frequency region of the Fourier spectrum of the data 
is dominated by instrumental frequencies such as $f_{\mathrm K}$,
$f_{\mathrm K}/2$, $f_{\mathrm Q}$ and $f_{\mathrm Q}/2$.
If we remove these instrumental peaks, we find the two largest 
non-instrumental ones: $f_{\mathrm B}$=0.01386~d$^{-1}$
and $f^{(1)}$=0.00819~d$^{-1}$. We obtain 
0.01394 and 0.00751~d$^{-1}$, respectively
for these values from the spacings of the side peaks. 
The two frequencies are in the ratio of $\approx1:2$
as we predicted. Since we detected sub-harmonic 
of the Blazhko frequency for several stars, $f^{(1)}$ could also be a 
sub-harmonic ($f^{(1)}=f_{\mathrm B}/2$) of $f_{\mathrm B}$.

The O$-$C diagram shows alternating
maxima and minima again (Fig.~\ref{o-c_all}). The Fourier spectrum
of this curve is much simpler than the spectrum of the original light curve data.
It contains $f_{\mathrm B}$=0.01395~d$^{-1}$ and its 
harmonic $2f_{\mathrm B}$, $f^{(1)}=f_{\mathrm S}=0.00761$~d$^{-1}$, and an additional
discrete peak close to 0 representing a global long-term period change.
Removing these three frequencies we obtain a residual spectrum which contains
two significant peaks at 0.01462~d$^{-1}$ (S/N=11.8) 
and 0.02010~d$^{-1}$ (S/N=4). The most plausible identification of
these peaks are $2f_{\mathrm S}$ and $f_{\mathrm B}+f_{\mathrm S}$, respectively.
These frequencies, especially the linear combination, contradict 
the sub-harmonic scenario. 

By removing all the mentioned frequencies
we receive a parabolic shape diagram (which can be seen in Fig.~\ref{o-c_all}).
The period decrease calculating from the best fitted parabola 
is $-8.4\times 10^{-9}$~dd$^{-1}$ that can not be explained by an evolutionary 
effect. It shows hidden period(s) and/or chaotic variation.  

To search for additional frequencies we 
pre-whitened the data with main frequency its harmonics and 
the largest 7-8 side peaks around the harmonics. 
We have found neither PD effect nor any
higher order radial overtone modes (Fig.~\ref{fr_add}).  

\subsubsection{V366\,Lyr = KIC~9578833}\label{v366lyr}

\begin{figure}
\includegraphics[width=8.5cm]{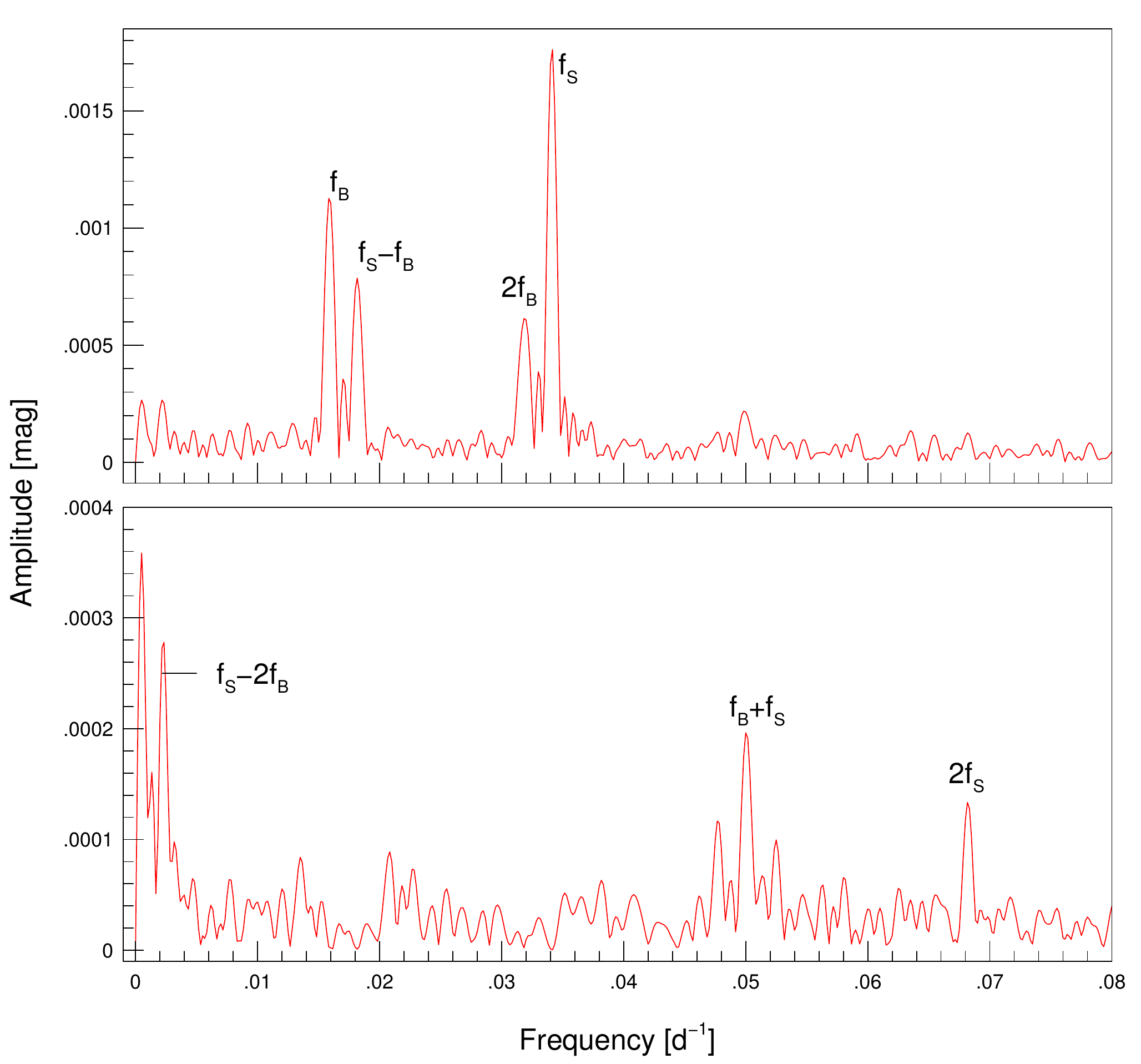}
\caption[]{
The Fourier spectrum of the O$-$C diagram of V366\,Lyr.
Top panel: a possible identification of the peaks.
Bottom panel: spectrum after we prewhitened with the 
four most significant frequencies above. 
} \label{V366_Lyr}
\end{figure}
The maxima of the light curve show a beating like phenomenon (Fig.~\ref{zoo}).
Calculations from the multiplet structures and the determination
of the significant low frequencies end in the same results.
Beside the primary modulation
frequency $f_{\mathrm B}$=0.0159~d$^{-1}$, two additional 
peaks can be detected at the frequencies 
$f^{(1)}=0.03415$ and $f^{(2)}=0.03175$~d$^{-1}$ with nearly equal amplitudes.
The latter is the first harmonic of the Blazhko 
frequency, but the first one seems to belong to a
secondary modulation frequency. Both of the primary  
($kf_0\pm f_{\mathrm B}$) and 
secondary modulation side peaks ($kf_0\pm f_{\mathrm S}$) 
 can be detected.
The ratio between the two modulation frequencies is close to 1:2.
 
The O$-$C diagram (in Fig.~\ref{o-c_all}) shows
a typical beating signal. 
The spectrum is surprising (Fig.~\ref{V366_Lyr}). 
Following our expectations it contains
two close frequencies at $f_{\mathrm B}$=0.015932~d$^{-1}$ 
and $f^{(3)}=0.01823$~d$^{-1}$ which must be responsible for
the beating phenomenon. Two other peaks are also visible.
One of them is the harmonic of the primary modulation 
frequency ($2f_{\mathrm B}$), while the other is at the 
position of $f^{(1)}=f_{\mathrm S}=0.03414$~d$^{-1}$. 
In this framework we can identify $f^{(3)}=f_{\mathrm S}-f_{\mathrm B}$.
Strangely enough, the amplitude of $f_{\mathrm S}$
is 1.55 times higher (1.7 vs 2.6~minutes) than that of $f_{\mathrm B}$.
In other words, the roles of the primary and secondary Blazhko 
modulations are reversed. This is the second case 
for this new phenomenon (see also V355\,Lyr in Sec.~\ref{v355lyr}). 

There is an alternate identification scenario for the detected 
frequencies. If we assume $f^{(3)}$ to be the genuine secondary modulation
then the amplitudes  would be in order:
$A(f_{\mathrm B})=1.7 > A(f_{\mathrm S})=1.1$~minutes,
however, $f^{(1)}$ should be identified as $f_{\mathrm B}+f_{\mathrm S}$.
In this case (i) the amplitude of the linear combination frequency
would be higher than any of its elements. Moreover, since
$f^{(1)}$ can be detected directly from the spectrum of the light curve,
(ii) \object{V366 Lyr} would be the only case where a linear combination 
frequency is detectable instead of the secondary modulation frequency.
For the above two (i and ii) reasons, we prefer the first scenario.

When we pre-whiten the O$-$C spectrum with the discussed four
frequencies,  additional significant peaks appear at some 
harmonics and linear combination frequencies, namely
at $2f_{\mathrm S}$, $f_{\mathrm S}+f_{\mathrm B}$  and
$2f_{\mathrm B}-kf_{\mathrm S}$ (bottom panel in Fig.~\ref{V366_Lyr}). 
After subtracting all these significant frequencies, 
the residual O$-$C diagram shows no more structures.

On the basis of Q1-Q2 data no additional 
frequency pattern has been found for V366\,Lyr \citep{Benko10}.
The situation is changed when we take into account the Q1-Q16 
time span. The highest peak between the harmonics is 
at the frequency $f^{(4)}=2.675799$~d$^{-1}$ (S/N=4.7, Fig.~\ref{fr_add}).
The period ratio is $P^{(4)}/P_0$=0.71. Frequencies with similar
period ratios were discovered in the CoRoT targets
\object{V1127 Aql} and \object{CoRoT 105288363}, and
 {\it Kepler} stars V445\,Lyr and \object{V360 Lyr} 
\citep{Chadid10, Guggenberger12, Benko10}. 
These frequencies are the dominant additional 
frequencies for only three stars: V1127\,Aql, V360\,Lyr and
V366\,Lyr. 

The referred papers generally explain these 
frequencies as excitation of independent non-radial modes. 
As we showed in \cite{Benko14} all of these
frequencies can be constructed by a linear
combination as $f^{(4)}=2(f_0-f_2)$. Here, in the case of V366\,Lyr,
we may identify the second overtone mode frequency 
with the marginal peak at $f_2=3.227711$~d$^{-1}$ (S/N=2.7).
This formal mathematical description has its 
own strengths and weaknesses. As opposed to the 
non-radial explanation, it could be verified or denied by
using existing radial hydrodynamic codes.
It is hardly understandable, however, why linear combination $2(f_0-f_2)$ 
is stronger than $f_0-f_2$.

\subsubsection{V360\,Lyr = KIC~9697825}\label{v360lyr}

\begin{figure}
\includegraphics[width=4cm,angle=-90,trim=100 30 100 100]{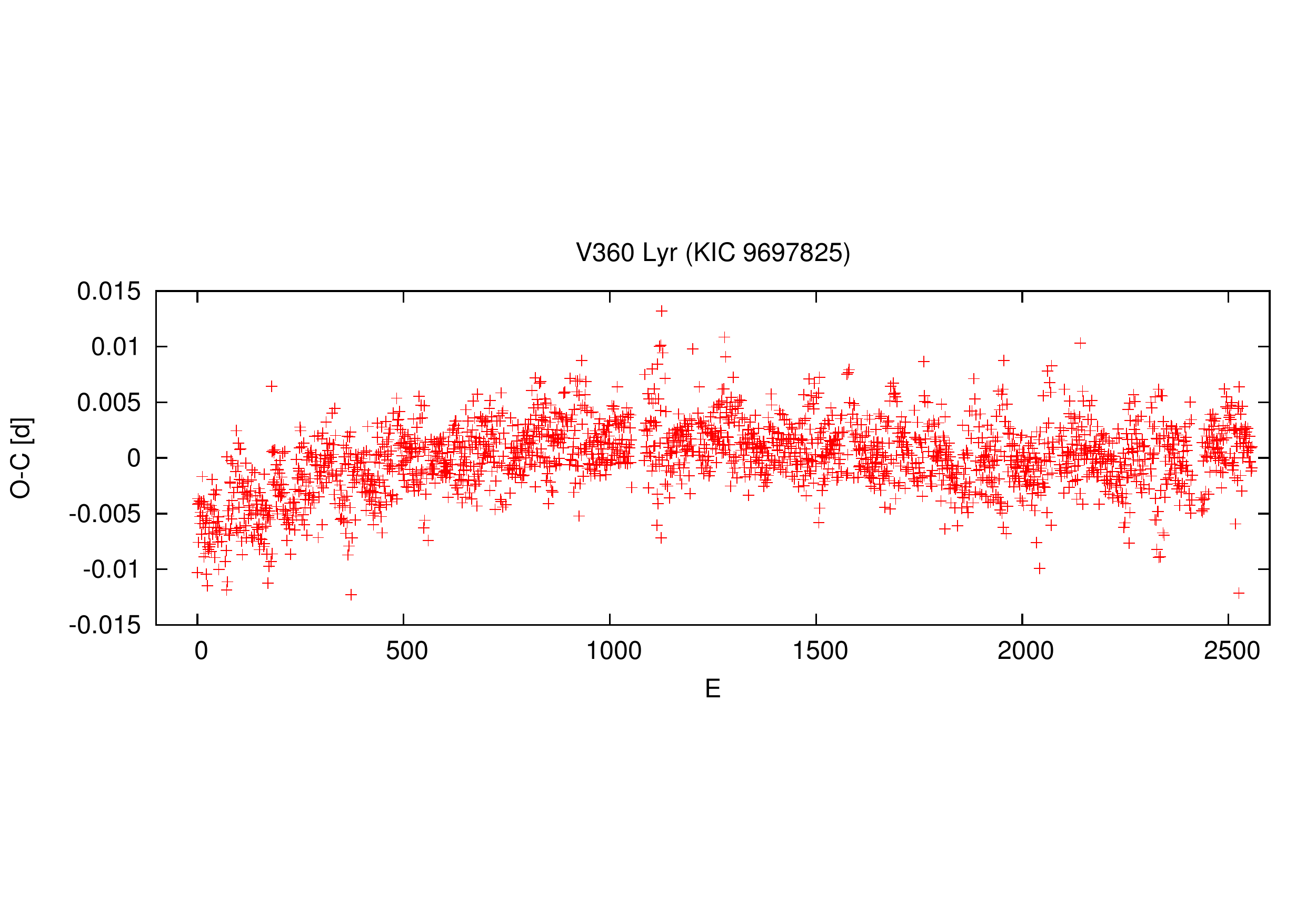}
\caption[]{
Residual O$-$C diagram of V360\,Lyr after we prewhitened
the data with all significant frequencies. 
} \label{9697825_o-c}
\end{figure}
The maxima of the light curve of V360~Lyr in Fig.~\ref{zoo}
show a slight beating phenomenon.
The Fourier spectrum contains rich multiplet structures around
the harmonics of the main pulsation frequency.
The largest side peaks indicate the following frequencies: 
$f_{\mathrm B}=0.01919$, 
$f^{(1)}=0.02374$, $f^{(2)}=0.02821$ and 
$f^{(3)}=0.04753$~d$^{-1}$. Two of them ($f_{\mathrm B}$ and $f^{(3)}$)
can be detected directly, as well. If we assume $f^{(3)}$ to be a
second independent modulation frequency ($f^{(3)}=f_{\mathrm S}$),
the other two values can be expressed as  
$f^{(2)}=f_{\mathrm S}-f_{\mathrm B}$ and $f^{(1)}=f_{\mathrm S}/2$.

Beyond the above mentioned four frequencies,
the Fourier spectrum of the O$-$C diagram shows two additional
peaks at $2f_{\mathrm B}$ and $f_{\mathrm S}+f_{\mathrm B}$. 
We found again a Blazhko star which shows not only a secondary
modulation but its half as well (see also discussion in Sec.~\ref{v355lyr}).
The ratio between the two modulation frequencies is 0.404
or 2:5. The residual curve of the
O$-$C diagram (Fig.~\ref{9697825_o-c}) shows a rather long
period oscillation as a secular period change. 

Two additional frequencies 
($f_1=2.4875$ and $f'=2.6395$~d$^{-1}$) of V360\,Lyr were
reported by \cite{Benko10} who explained $f_1$
with a possible first overtone mode and $f'$ 
with an independent non-radial mode.  
As \cite{Szabo10} have already mentioned 
$f'$ could also be a member of a PD pattern around $f'=1.5f_0$.
When we analyze the Q1-Q16 data set the highest
amplitude peak in this region is located at $f^{(4)}=2.678669$~d$^{-1}$
($P^{(4)}/P_0=1.49$) which is a PD frequency without doubt.

Interpretation of the strongest additional peak at the 
frequency of $f^{(5)}=2.487740$~d$^{-1}$ is more problematic.
The period ratio $P^{(5)}/P_0=0.721$ is far from the 
canonical value of the first radial overtone and fundamental modes (0.745). 
Such a ratio could be produced 
by a highly metal abundant RR\,Lyrae star (see e.g. fig.~8 in \citealt{Chadid10})
but the metallicity of V360\,Lyr is [Fe/H]=$-1.5\pm 0.35$~dex (N13).
This difference might also be
due to the resonance, which excites this mode
in non-traditional way \citep{Molnar12}. 
We suggested in \cite{Benko14} an other explanation:
similarly to V366\,Lyr (and V445\,Lyr), this frequency 
could be a linear combination $2(f_2-f_0)$, where 
$f_2=3.036015$~d$^{-1}$ is the frequency of the second radial overtone mode.

\subsubsection{KIC~9973633}\label{kic9973633}

The history of the star is the same as that of KIC~7257008:
the ASAS survey discovered it and its basic parameters
were determined for the first time by N13.
The {\it Kepler} data set of KIC~9973633 is relatively the
most unfortunate in the analyzed sample. It has short 
time coverage (data exist from only Q10) and additional two  
quarters (Q11 and Q15) are missing (Fig.~\ref{zoo}). 

Although the triplet structures around the harmonics
provide us the Blazhko frequency $f_{\mathrm B}$=0.01490~d$^{-1}$, we 
can not detect it directly in the low frequency range of the
Fourier spectrum. Here the instrumental peak $f_{\mathrm K}$
dominates. When we remove it, we find two high amplitude peaks at 
$f_{\mathrm K}/2$ and $f^{(1)}=0.00411$~d$^{-1} = f_{\mathrm B}-f_{\mathrm Q}$.  
After the next pre-whitening step we can see three
peaks near the detection threshold at $f_{\mathrm Q}$,
$f^{(2)}=0.03771$~d$^{-1}$ and $f^{(3)}=0.02701$~d$^{-1}$.
It is easily recognizable that $f^{(3)}=f^{(2)}-f_{\mathrm Q}$. 
Which one is the real independent frequency:
 $f^{(2)}$ or $f^{(3)}$?

To investigate the question we analyzed the O$-$C diagram of
KIC~9973633. The Fourier spectrum
shows two highly significant peaks at $f_{\mathrm B}=0.01486$~d$^{-1}$
and $2f_{\mathrm B}=0.02972$~d$^{-1}$ (Fig.~\ref{fr_low}). By pre-whitening the data with 
these two frequencies we obtain a third one at 0.03675~d$^{-1}$.
If we identify it with $f^{(3)}=f_{\mathrm S}$, we are in a similar situation
as in the case of V360\,Lyr (Sec.~\ref{v360lyr}): we have two modulation frequencies
with the ratio of 2:5. 
The residual of the O$-$C diagram does not show any structures:
it is a constant line with some scatter.

\subsubsection{V838\,Cyg = KIC~10789273}\label{v838cyg}

\begin{figure}
\includegraphics[width=9cm]{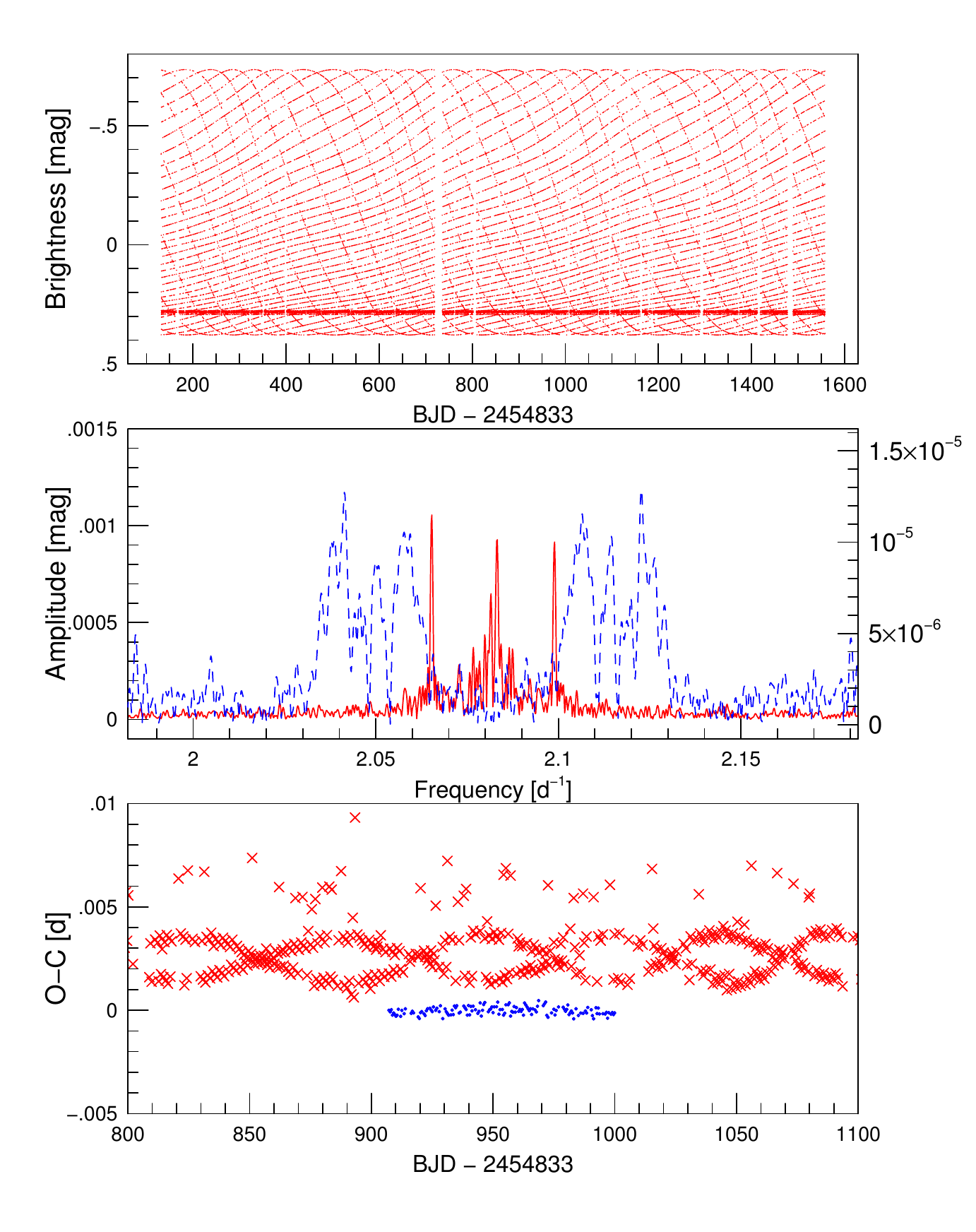}
\caption[]{
Analysis of \object{V838 Cyg}. The top panel shows the synthetic light
curve prepared by using the Fourier parameters of the observed
data without assuming any modulations and sampled on the 
{\it Kepler} data points (cf. Fig.~\ref{zoo}). Middle panel: Fourier spectra
of the observed (red continuous line) and synthetic (blue dashed line)
time series around the pulsation frequency after the data were 
pre-whitened with the main frequency. The left and right hand side magnitude scales belong
to the observed  and synthetic spectra, respectively.  Bottom panel: 
Part of the O$-$C diagrams for LC data (red crosses) and SC observations
(blue dots). 
} \label{V838_Cyg}
\end{figure}
The extremely low amplitude 
Blazhko modulation of V838\,Cyg was discovered by N13.
The light curve in Fig.~\ref{zoo} shows wavy maxima and minima, however,
this is structure produced by interference of
the sampling and the pulsation frequencies and it does not indicate any AM.
It is a virtual modulation and not a real one.
It means that we need more careful investigations to decide whether 
V838\,Cyg is modulated or not.
Clear triplets around the main pulsation frequency  
and its harmonics in the Fourier spectrum suggest a modulation 
phenomenon with the frequency of $f_{\mathrm B}=0.01681$~d$^{-1}$. 
It follows that the period is $P_{\mathrm B}=59.5$~d which roughly 
agrees with the dominant modulation period value (54-55~d) 
found by N13. 

V838\,Cyg was observed in SC mode 
in only one quarter in Q10. We processed these pixel data 
in the same way as we have done for LC data.
The obtained SC light curve does not show virtual modulation
like LC data, but shows a small amplitude change. This variation is, however, hard to distinguish
from possible instrumental trends. To test the presence of 
the modulation we made synthetic data. We prepared an artificial time series 
$m_{\mathrm{syn}}(t)$ using the Fourier parameters of the observed data such as
$f_0$, $A_k$ and $\varphi_k$, where $A_k$ and $\varphi_k$ mean
the amplitude and phase values of the $k$th harmonics ($k=1, 2,\dots,11$)
i.e. without modulation side peaks:
\[
m_{\mathrm{syn}}(t)=\sum_{k=1}^{11}A_k \sin (2\pi f_0 t + \varphi_k).
\]
The synthetic data were sampled at the observed time stamps $t$
(top panel in Fig.~\ref{V838_Cyg}).

The spectra of the observed and synthetic
data have systematic differences (see middle panel in Fig.~\ref{V838_Cyg}). 
(i) In the case of synthetic
data the side peaks have $\approx80$ times smaller 
amplitudes than the observed ones (0.012~mmag vs. 1~mmag). (ii) The position of the side
frequencies are different. (iii) The structures of these
side patterns are also different: the observed data show 
simple clear triplet, while the synthetic data produce complicated
multiplets. To summarize, we confirmed 
the AM with an independent method.
Since the amplitude of the modulation
is very small, we could not detect $f_{\mathrm B}$ directly from
the low frequency range of the Fourier spectrum. This part of the 
spectrum is dominated by instrumental frequencies 
($f_{\mathrm K}$, $f_{\mathrm L}$, where $P_{\mathrm L}=1/f_{\mathrm L}$ indicates 
the total length of the observation: $\sim4$~years). 
N13 mentioned possible multiple modulations. 
After pre-whitening the data with the side frequencies 
$kf_0\pm f_{\mathrm B}$ only instrumental side peaks 
($kf_0\pm f_{\mathrm K}$ $kf_0\pm f_{\mathrm L}$) remain 
significant around the harmonics.    

We can not construct the O$-$C diagram the traditional
way because the sparse continuous sampling with the relatively short
pulsation period and the interpolation errors produce 
systematic undulations (in the bottom panel in Fig.~\ref{V838_Cyg}). 
Due to the long Blazhko period, the analysis of SC data does not help too much.
The sensitive method used by N13 resulted in 
0.001~radian phase variation in $\varphi_1$ which predicts about
$5\times 10^{-4}\approx 40$~s  O$-$C variation. It is about
our detection limit: the standard deviation of O$-$C for SC data 
is $\approx 2\times 10^{-4}$. Our analysis supports 
the phase modulation reported by the discovery paper.

We also found additional frequency patterns for the first time in this star. 
After we subtracted the significant harmonics of the 
pulsation frequency and triplets around them, 
the following additional frequencies can be detected
between $f_0$ and $2f_0$:
$f_2=3.509142$~d$^{-1}$, $1.5f_0=3.113906$~d$^{-1}$ (PD) 
and $3f_0-f_2=2.737256$~d$^{-1}$, respectively
(see Fig.~\ref{fr_add}). 

\subsubsection{KIC~11125706}\label{kic11125706}

Though KIC~11125706 shows the second lowest amplitude AM
in our sample, its long pulsation period allows us to
detect it without any problems. The asymmetric triplet structures 
around the pulsation frequency and its harmonics provide
$f_{\mathrm B}$. Direct detection of this frequency has failed.
The low frequency range of the Fourier spectrum is dominated by 
instrumental peaks. 

Contrary to this, the Fourier spectrum of the O$-$C diagram 
(Fig.~\ref{fr_low}) contains a very significant (S/N=37.8) 
peak at $f_{\mathrm B}$.
Pre-whitening with this frequency, a lower amplitude,
but clearly detectable ($A(f^{(1)})=1.9\times 10^{-4}$~d, S/N=$6.4$) 
peak can be seen at $f^{(1)}=0.01698$~d$^{-1}$. 
Is it a real secondary modulation frequency or results from the
sampling? We generated synthetic O$-$C data using the Fourier 
parameters of the primary Blazhko period and 
added Gaussian noise to it. The synthetic diagram was sampled 
exactly at the same points as the observed one. The Fourier 
spectrum of the artificial O$-$C curve contains only the
$f_{\mathrm B}$ and it does not show any other significant peaks.
Thus we ruled out an instrumental origin of this frequency.
Now, we tend to identify $f^{(1)}=f_{\mathrm S}$. In this case 
the two modulation frequencies are in a ratio of nearly 3:2.
The quadratic fit of the residual O$-$C gives a slight period increase:
$dP_0/dt=4.4\times 10^{-10}\pm1.1\times10^{-10}$~dd$^{-1}$. 

Returning to the light curve analysis we do not 
find any significant frequencies around $f_{\mathrm S}$ 
and no side peaks detected which could belong to this
frequency. These facts mean that the AM of $f_{\mathrm S}$
is below of our detection limit. Furthermore, no additional peak 
patterns have been found between harmonics (Fig.~\ref{fr_add}).

\subsubsection{V1104\,Cyg = KIC~12155928}\label{v1104cyg}

Beyond V2178\,Cyg, V1104\,Cyg was also the subject of
the case study of N13. We can add only 
few things to this study. We derived the Blazhko cycle based on a longer time span.
The frequency of the Blazhko modulation $f_{\mathrm B}$ is
highly significant both in the Fourier spectrum of the light
curve and that of the O$-$C diagram. The latter spectrum also includes the 
harmonic $2f_{\mathrm B}$ (Fig.~\ref{fr_low}) 
indicating non-sinusoidal behavior of the
FM. The O$-$C residual does not indicate any
period changes during the {\it Kepler} time-scale.
Our analysis resulted in neither secondary
modulation nor additional frequencies between harmonics (Fig.~\ref{fr_add}).

\section{Conclusions}

The main goal of this paper was to investigate 
the long time-scale behavior of the Blazhko effect 
among the {\it Kepler} RR\,Lyrae stars. 
To provide the best input for the analysis we prepared 
time series from the pixel photometric data with our own tailor-made aperture.
These light curves include the total flux of the 
stars for only 9 cases while some portion of the flux of 6 RR\,Lyrae stars
was lost using even in the largest possible aperture. 
Nevertheless, our data set comprises the longest continuous, 
most precise observations of Blazhko RR Lyrae stars
ever published. These data will be unprecedented for years 
to come.

Since the Blazhko effect manifests itself in simultaneous
AM and FM we analyzed both phenomena
and compared the results of the separated analyses.
This approach reduces the influence of the instrumental effects.

We detected single Blazhko period for three stars:
V783\,Cyg, V838\,Cyg and V1104\,Cyg. Since
V838\,Cyg shows the smallest amplitude modulation both in AM
and FM, we could not confirmed
its multiperiodicity which was suspected by N13. 

Twelve stars in our sample show evidences for multiperiodic
modulation. In eight cases we could determine two significant Blazhko
periods, while for four additional cases 
(V2178\,Cyg, V808\,Cyg, V354\,Lyr and KIC~7257008) we could 
establish the presence of a possible long secondary period.
It does not mean, however, that we could described the total 
variations with these one or two periods completely. The residual curves show 
significant structures (see e.g. V2178\,Cyg, V808\,Cyg, V450\,Lyr)
after subtracting the best fitted light/O$-$C curves.
For the 2-3 stars with the shortest Blazhko periods we have a
chance to carrying out a dynamical analysis. In this work we
mentioned the preliminary result from the cycle-to-cyle variation of
V783\,Cyg modulation (Plachy et al. in prep.) which hints for its chaotic nature.

The latest and most complete 
compilation of the Galactic Blazhko RR\,Lyrae stars \citep{Skarka13}
consists only 8 multiperiodic cases among 242 known field Blazhko 
RR\,Lyrae stars (3.3\%). More recently \cite{Skarka14} studied 
a more homogeneous 321 elements sample from the ASAS and SuperWASP surveys.
He found the ratio of the multiperiodic and irregularly modulated stars
to be 12\%. In this work 
we surprisingly found that most of the {\it Kepler} Blazhko stars
-- 12 from 15 (80\%) -- belongs to the multiperiodic group.
 In other words: the Blazhko effect predominantly manifests 
as a multiperiodic phenomenon instead of a 
mono-periodic and regular one.
Here we briefly summarize the main characteristics of the
multiperiodic modulations. 

\begin{itemize}
\item
Up to now the known smaller amplitude (secondary) modulation 
periods were generally longer than the primary ones.
The only exception is RZ\,Lyr \citep{Jurcsik12}. 
Here we showed five further examples 
(V355\,Lyr, V450\,Lyr, V366\,Lyr, V360\,Lyr and
KIC~9973633) for shorter secondary periods.

\item
What is more, the definition of the primary and secondary modulations
proved to be relative. In three cases
(for V450\,Lyr, V366\,Lyr and V355\,Lyr) the relative strength 
of the primary modulations is weaker in FM than in AM.

\item
The linear combination of the modulation frequencies can
generally be detected that indicates the nonlinear coupling between the
modes. Sub-harmonic frequencies ($f_{\mathrm B}/2$ and/or $f_{\mathrm S}/2$)
were detected for numerous cases (KIC~7257008, V355\,Lyr, V450\,Lyr, V360\,Lyr).
In the majority of the cases
the two modulation periods are in a ratio of small integer 
numbers:
1:2 for V353\,Lyr; 2:1 for V366\,Lyr and V355\,Lyr; 
(3:2 for V2178\,Cyg); 2:3 for KIC~11125706;
5:2 for KIC~9973633 and V360\,Lyr. (Here the ratios of the 
primary AM period vs. secondary one are indicated.) 
\end{itemize}

As a by-product of the analysis we report here for the 
first time additional frequency structures for V808\,Cyg,
V355\,Lyr and V838\,Cyg. For all three cases we detected
the second radial overtone mode $f_2$. 

The former studies of {\it CoRoT} and {\it Kepler} Blazhko data 
found unidentified frequencies for numerous stars. 
These frequencies were explained by non-radial mode excitation.
Here we showed that almost all of such frequencies can also
be produced by linear combinations of radial modes.
The only case where we could not find a proper linear combination is
the highest amplitude additional frequency of V354\,Lyr. 

The amplitudes of these frequencies point to rather the non-radial 
mode scenario. These non-radial modes may be excited by resonances at the locations of
the linear combination of the radial modes \citep{VanHolst}.
In other words, these frequencies are linear combinations from mathematical 
point of view only, and physically they are frequencies of independent 
(non-radial) modes.

\

\acknowledgments{
Funding for this Discovery mission is provided by NASA's Science Mission Directorate. 
This project has been supported by 
the `Lend\"ulet-2009 Young Researchers' Program of 
the Hungarian Academy of Sciences, the Hungarian OTKA grant K83790 and
the KTIA URKUT\_10-1-2011-0019 grant. 
The research leading to these results has received funding from the
European Community's Seventh Framework Programme (FP7/2007-2013) under
grant agreements no. 269194 (IRSES/ASK) and  no. 312844 (SPACEINN).
The work of E. Plachy was supported by the European Union and 
the State of Hungary, co-financed by the European Social Fund in the framework 
of T\'AMOP 4.2.4.\ A/2-11-1-2012-0001 `National Excellence Program'.
R. Szab\'o was supported by the J\'anos Bolyai Research Scholarship of the 
Hungarian Academy of Sciences. 
The authors thank the referee for carefully reading  our 
manuscript and for his/her helpful suggestions.
}

\end{document}